\documentclass{aa}
\usepackage{graphicx}
\usepackage{natbib}

\begin{document}
   \title{The sub-arcsecond dusty environment of Eta Carinae\thanks{Based on observations
 collected at the European Southern Observatory, Chile}}

\titlerunning{The sub-arcsecond dusty environment of Eta Carinae}
 %  \subtitle{I. Spherical geometries}

   \author{O. Chesneau
          \inst{1}              \and
         M.~Min \inst{2}
          \and
          T.~Herbst
          \inst{1}              \and
          L.B.F.M.~Waters
        \inst{2}
            \and
            D.J.~Hillier
            \inst{3}
          \and
          Ch.~Leinert
            \inst{1}
            \and
        A.~de~Koter
            \inst{2}
            \and
            I.~Pascucci
            \inst{1}
            \and
           W.~Jaffe
           \inst{4}
           \and
           R.~K\"{o}hler
           \inst{1}
            \and
            C.~Alvarez
\inst{1}
 \and
 R.~van~Boekel
\inst{2}
\and
            W.~Brandner
\inst{1}
            \and
            U.~Graser
            \inst{1}
\and
 A.M.~Lagrange
\inst{5}
 \and
            R.~Lenzen
\inst{1}
\and
S. Morel
\inst{6}
\and
            M.~Sch\"{o}ller
\inst{6}
     }

   \offprints{O. Chesneau}

   \institute{Max-Planck-Institut f\"{u}r Astronomie,
   K\"{o}nigstuhl 17, D-69117 Heidelberg, Germany\\
              \email{chesneau@mpia-hd.mpg.de}
                      \and
                      Sterrenkundig Instituut `Anton
Pannekoek', Kruislaan 403, 1098 SJ Amsterdam, The Netherlands \and
Department of Physics and Astronomy, University of Pittsburgh,
3941 O'Hara Street, Pittsburgh, PA 15260, USA \and Leiden
Observatory, Niels Bohr weg 2, 2333 CA Leiden, The Netherlands,
\and Laboratoire d'Astrophysique de l'Observatoire de Grenoble,
Universit\'{e} J. Fourier, CNRS, BP 53, 38041 Grenoble Cedex 9,
France \and European Southern Observatory, Casilla 19001,
Santiago, Chile }

   \date{Received; accepted }

   \abstract{The core of the nebula surrounding
Eta Carinae has been observed with the VLT Adaptive Optics system
NACO and with the interferometer VLTI/MIDI to constrain spatially
{\it and} spectrally the warm dusty environment and the central
object. In particular, narrow-band images at 3.74~$\mu$m and
4.05~$\mu$m reveal the butterfly shaped dusty environment close to
the central star with unprecedented spatial resolution. A void
whose radius corresponds to the expected sublimation radius has
been discovered around the central source. Fringes have been
obtained in the Mid-IR which reveal a correlated flux of about
100~Jy situated 0$\farcs$3 south-east of the photocenter of the
nebula at 8.7~$\mu$m, which corresponds with the location of the
star as seen in other wavelengths. This correlated flux is partly
attributed to the central object, and these observations provide
an upper limit for the SED of the central source from 2.2~$\mu$m
to 13.5~$\mu$m. Moreover, we have been able to spectrally disperse
the signal from the nebula itself at PA=318 degree, i.e. in the
direction of the bipolar nebula ($\sim$310$^\circ$) within the
MIDI field of view of 3$\arcsec$. A large amount of corundum
(Al$_2$O$_3$) is discovered, peaking at 0$\farcs$6-1$\farcs$2
south-east from the star, whereas the dust content of the Weigelt
blobs is dominated by silicates. We discuss the mechanisms of dust
formation which are closely related to the geometry of this
Butterfly nebulae.

   }

   \maketitle
%
%________________________________________________________________

\section{Introduction}
Eta Carinae is one of the best studied but least understood
massive stars in our galaxy (Davidson and Humphreys, 1997). With a
luminosity of 5$\times$10$^6$ L$_\odot$, it is one of the most
luminous stars and at 10~$\mu$m it is one of the brightest objects
outside the solar system (Neugebauer \& Westphal, 1969). Eta Car
is classified as a Luminous Blue Variable (LBV). Among the most
prominent characteristics of the unstable LBV phase are strong
stellar winds and possible giant eruptions, which lead to the
peeling off of the outer layers of the H-rich stellar envelope and
the formation of small ($\sim$0.2-2~pc) circumstellar nebulae
(Nota et al. 1995).

During the last two centuries, Eta Carinae has lived through a
turbulent history. During the great eruption in the 1840s, the
large bipolar nebula surrounding the central object (known as the
``Homunculus'') was formed. Currently, the Homunculus lobes span a
bit less than 20$"$ on the sky (or 45000 AU at the system distance
of 2.3 kpc) and they are largely responsible for the huge infrared
luminosity of the system. The cause of the outburst remains
unknown. The chemical composition of the Homunculus gas is not
known, but some studies of the ionized outer ejecta (Lamers et al.
1998, Smith \& Morse 2004 and references herein) suggest an
overabundance of N and a severe depletion of C and O. Such an
abundance pattern is consistent with CNO equilibrium burning, and
suggests a highly evolved star at the base of the ``explosion'' .

The central source was studied by speckle interferometry
techniques, which revealed` a complex knotty structure (Weigelt \&
Ebersberger 1986; Falcke et al.\ 1996). At first, three remarkably
compact objects between 0$\farcs$1 and 0$\farcs$3 northwest of the
star were isolated (the so-called BCD {\it Weigelt blobs}, the
blob A being the star itself). Other similar but fainter objects
have been detected since then (Weigelt et al 1995, 1996; Davidson
et al 1997). It has been found that they are surprisingly bright
ejecta moving at low speeds ($\sim$50 km.s$^{-1}$) and that they
belong to the equatorial regions close to the star. Their
separation from the star is typically 800 AU. The large scale
equatorial midplane debris-disc was nicely revealed with HST data
(Morse et al. 1998).

The detection of a 5.52 year period in spectroscopy and
near-infrared photometry (Damineli 1996) in Eta Carinae has been
confirmed (Damineli et al.\ 2000 Davidson et al.\ 2000, Abraham et
al. 2003, Corcoran 2003, Whitelock et al. 2003, 2004). The
existence, mass and orbit of a companion and its impact on the
behavior of the primary are still strongly disputed (e.g. Davidson
1999, Duncan et al. 1999; Stevens \& Pittard 1999; Corcoran et al.
2001, Feast et al. 2001, Pittard \& Corcoran 2002; Duncan \& White
2003).

Dust plays a key role in the study of Eta Car. It intervenes in
every observation as strong and patchy extinction. It also allows
the mass of the nebula to be determined. Dust has also been
frequently invoked as an important process in  explaining the
photometric variability of Eta Car. However, the exact influence
and location of dust formation/destruction has never been observed
and established. Eta Car was observed with the Infrared Space
Observatory Observatory (ISO) by Morris et al.\ (1999). The ISO
spectra indicated that a much larger amount of matter should be
present around Eta Car in the form of cold dust than previously
estimated. Observations with higher spatial resolution by Smith et
al.\ 2002a and Smith et al.\ 2003a showed a complex but organized
dusty structure within the three inner arcseconds. They showed
that the dust content around the star is relatively limited and
claimed that the two polar lobes should contain a large mass of
relatively cool dust which could explain the ISO observations.

The high spatial resolution images of the equatorial regions are
puzzling in several ways and raise new key questions in addition
to the numerous ones still to be answered: Why was the
eruption azimuthally asymmetric? Is the complex geometry of the
dusty torus a consequence of the 1840 outburst or has it been
affected by more recent events? What is the status of the complex
Weigelt blob region?  Why did the star eject such slow-moving
material in its equatorial zone? Does the star form dust continuously,
or in  episodes related to the mini outburst or the putative
wind-wind interaction of Eta Car with its companion?

Improved spatial resolution is a major reason for the recent
progress in our study of this emblematic star. HST STIS
observations allowed the observation of the stellar spectrum of
Eta Car roughly separated from its nearby ejecta (Hillier et al.
2001). Moreover this impressive instrument has allowed the study
of the stellar wind from several latitude points of view in the
nebulae by means of P Cygni absorption in Balmer lines reflected
in the nebula (Smith et al. 2003). The asphericity of the wind
properties has been convincingly proven and Smith et al. suggested
that the observed enhanced polar wind mass-loss rate may be
explained through the theoretical frame developed by Stan Owocki
and collaborators (Dwarkadas \& Owocki 2002). In their model, an
enhanced mass loss occurs along the rotation axis, due to the
large temperature difference between pole and equator, caused in
turn by the rapid rotation of the star (the von Zeipel effect).
Recently, the ionized stellar wind of Eta Carinae has been
resolved on the 5 milliarcsecond (mas) scale at a wavelength of
2.2~$\mu$m with data obtained with VINCI on the Very Large
Telescope Interferometer (VLTI, van Boekel et al.\ 2003). These
observations are consistent with the presence of one star which
has an ionized, moderately clumpy stellar wind with a mass loss
rate of about 1.6x10$^{-3}$ M$_{\odot}$/yr. This star-plus-wind
spherical model, developed by Hillier et al. (2001), is also
consistent with the HST STIS observations of the central object.
It has also been found that the star is elongated with a
de-projected axis ratio of about 1.5 and that the axis itself is
aligned with the axis of the large bipolar nebula. These VLTI
observations gave an important confirmation of the wind geometry
previously proposed by Smith et al. 2003.

The Hillier model suggests a flux level of 200-300~Jy at 10~$\mu$m
and a spatial extent of the star plus wind of 10-15 mas diameter.
These dimensions can only be probed using MIDI at the VLTI. The
VINCI observations do not require the presence of other components
in the core. In particular, no evidence for a hot dust disc or the
putative companion were found. However the presence of warm
(300-600~K) dust in the immediate surroundings of  Eta Car cannot
be excluded since it is too cool to be detected at 2~$\mu$m.

The MIDI recombiner attached to the VLTI is the only instrument
that is able to provide  sufficient spatial and spectral
resolution in the mid-infrared to disentangle the central
components in the Eta Car system from the dusty environment. By
definition an interferometer measures a correlated flux, i.e. a
flux originating from a sufficiently spatially unresolved source
so that it is able to produce fringes. The measured correlated
flux depends on the source total flux, its geometry and on the
length and direction of the projected baseline(s) of the
interferometer. We used the 102~m baseline between the telescopes
Antu (UT1) and Melipal (UT3) to observe, for the first time, Eta Car
with a resolution of 5-10~mas through the entire N band.

These observations have been complemented with broad- and
narrow-band observations taken with the NAOS/CONICA (NACO) imager
installed on UT4 (Kueyen), equipped with an adaptive optics (AO)
system. The diffraction limit at 3.8~$\mu$m is about 100~mas. At
this wavelength, the NACO adaptive optics is less constrained by
the atmosphere, providing routinely an excellent correction
(Strehl ratio reaching 0.5). A careful deconvolution procedure can
improve it to about 50-80 mas. The NACO observations offer the
opportunity to bridge the gap to the high resolution data of MIDI
obtained with a sparse UV coverage.

In Section \ref{sec:obs}, we describe the observations and the
data reduction. We analyze the NACO and MIDI images in
Section~\ref{sec:images}, and then we examine the spatial
distribution of the dust close to the star in
Section~\ref{sec:dust}. The information extracted from the
correlated flux detected by MIDI is presented in
Section~\ref{sec:correlated}. Finally in
Section~\ref{sec:discussion}, we summarize the implications of the
extracted information.

\section{Observations and data reduction}
\label{sec:obs}
\subsection{NACO high resolution imaging}
We have observed Eta Carinae with the adaptive optics camera NACO
(Lenzen et al. 1998; Rousset et al. 2003) attached to the fourth
8.2 m Unit Telescope of the Very Large Telescope (VLT) of the
European Southern Observatory (ESO), located at Cerro Paranal,
Chile. NAOS was operated in the visual wavefront sensor
configuration with the SBRC Aladdin 1024x1024 detector. We
observed with J, H, Ks with the S13 camera and L$'$ broad-band
filters and the NB\_374, NB\_405 narrow band filters that cover
the emission lines Pfund~$\gamma$ and Bracket~$\alpha$
respectively with the L27 camera. Using camera mode S13 and L27,
the field of view was 14$\arcsec$x14$\arcsec$ and
28$\arcsec$x28$\arcsec$ respectively and the pixel scale was 13.25
and 27.1~mas per pixel respectively, a size sufficient to satisfy
the Nyquist sampling criterion. 13.25~mas and 27.1~mas correspond
to 30 and 62~AU respectively at the distance of 2.3~kpc. The
AutoJitter mode was used, that is, at each exposure, the telescope
moves according to a random pattern in a 10$\arcsec$ box.
Cross-correlation was used to recenter the images at about 0.15
pixel accuracy.

\begin{table}[h]
 \caption{\label{tab:NACO}Journal of observations with NACO/UT4. The phase within the 5.52-year cycle
 is computed from the ephemeris of Daminelli et al. (2000).}
\vspace{0.3cm}
\begin{center}
 \begin{tabular}{lllcc}
 \hline
 \hline
 Star & Filter & Camera & Time &$t_{\rm exp}$ \\
  \hline
\multicolumn{5}{c}{15/16-11-2002, JD=2452625, $\Phi=0.87$}\\
 \hline
$\eta$ Car & J & S13&T08:12:00&86s\\
$\eta$ Car & H & S13&T08:16:02&86s\\
$\eta$ Car & Ks &S13&T08:04:21&86s\\
$\eta$ Car & L$'$ &L27&T08:20:24&50s\\

 \hline
\multicolumn{5}{c}{16/17-11-2002, JD=2452626, $\Phi=0.87$}\\
 \hline
$\eta$ Car & NB\_374 & L27&T08:47:00&50s\\
$\eta$ Car & NB\_405 & L27&T08:50:44&50s\\
HD 101104 & NB\_374 &L27&T09:10:17&50s\\
HD 101104 & NB\_405 &L27&T09:17:51&50s\\

 \hline
  \end{tabular}
\end{center}
 \end{table}
A neutral density filter with an attenuation factor of 70 was used
in order to avoid saturating the central peak of the point spread
function. However, the L$'$ image was saturated within the first
0$\farcs$5 even using the shortest exposure time possible (0.17s).
The NB\_405 narrow-band image is not saturated but the peak of the
central source is in the non-linearity regime of the detector. The
NB\_374 narrow-band image does not suffer from this effect due to
the lower continuum and line fluxes at this wavelength and the
slightly narrower filter.

\begin{figure*}
  \begin{center}
    \includegraphics[height=8.6cm]{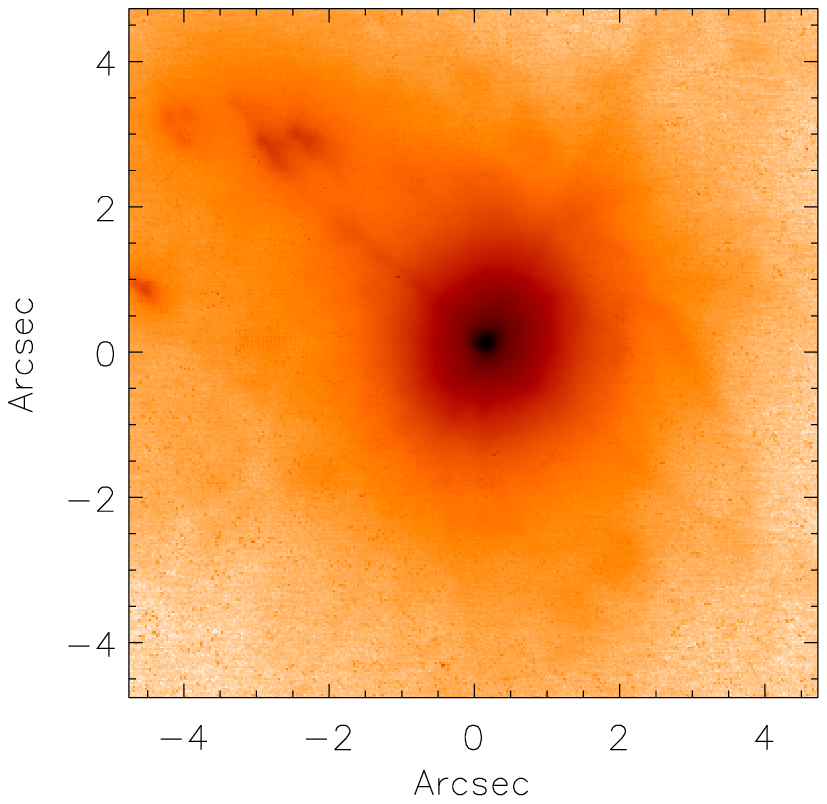}
\includegraphics[height=8.6cm]{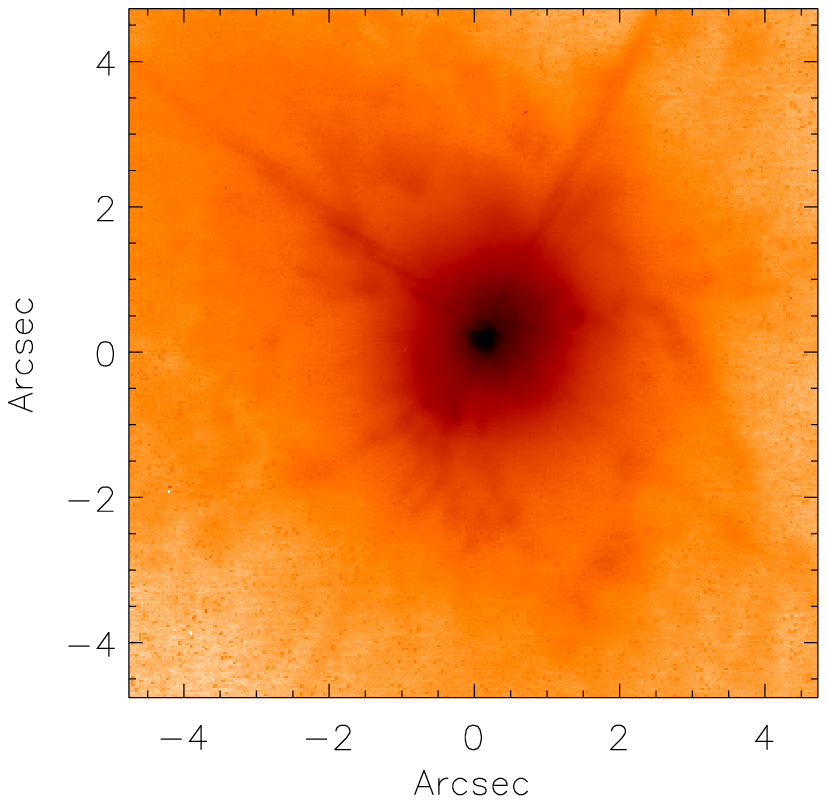}
    \includegraphics[height=8.6cm]{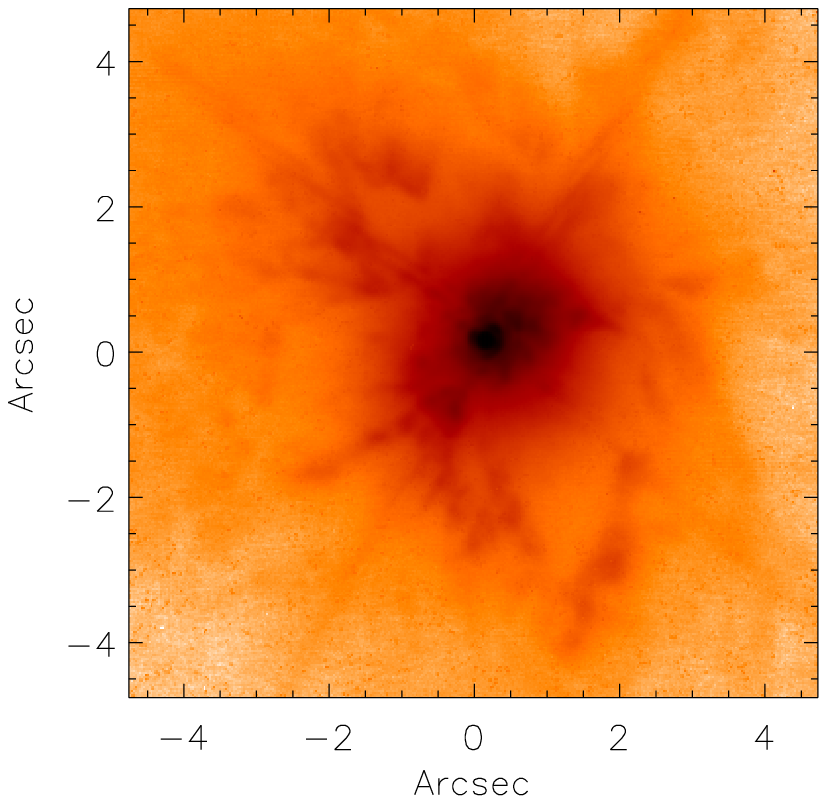}
\includegraphics[height=8.6cm]{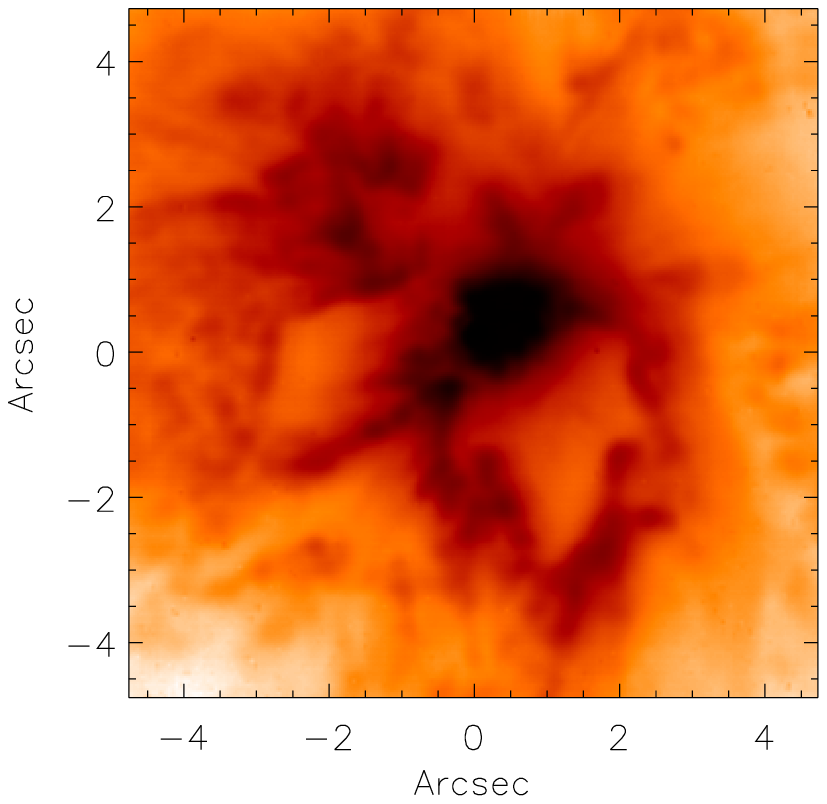}

  \end{center}
  \caption[]{\label{fig:continuum}From left to right and up to down, J, H, Ks, L$'$ images from NACO
  shown in logarithmic inverted scale. The L$'$ image is slightly overexposed in spite of the smallest
  possible integration time. }
\end{figure*}

\begin{figure*}
  \begin{center}
     \includegraphics[height=14.cm]{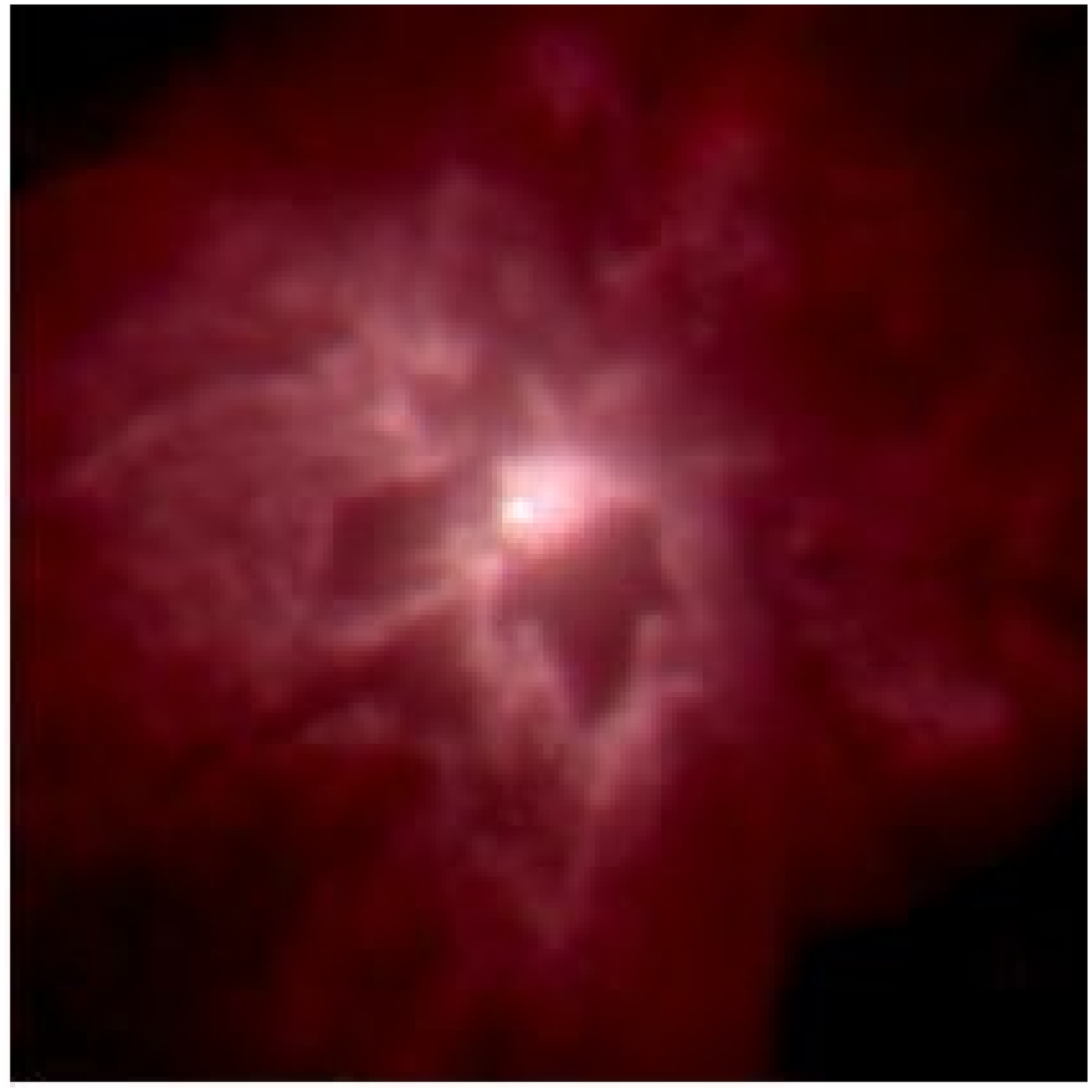}

  \end{center}
  \caption[]{\label{fig:color}Color image combining the flux from L$'$ (red) and Br$\alpha$ (green)
  and Pf$\gamma$ (blue) filters. At these
  wavelengths, the hydrogen emission represents only a small amount
  of the flux: about 10\% and 2\% of flux recorded with the Br$\alpha$ and
Pf$\gamma$ filter respectively. The L$'$ filter shows deeper
details of the nebula but is saturated in the
0$\farcs$5x0$\farcs$5 core. The white bar represents 1$\arcsec$.
  }
\end{figure*}

Individual dithered exposures were co-added, resulting in a total
exposure time $t_{\rm exp}$ shown in Table~\ref{tab:NACO}. The
data reduction has been performed using a self-developed IDL
routine that processes the individual frames as follows. First,
bad pixels are removed. Then, the sky is
computed as the mean of the dithered exposures, and subtracted
frame by frame. Finally, all the sky-subtracted frames are shifted
and added together. The reduced broad-band images are shown in
Fig.~\ref{fig:continuum}. These broad-band images have not been
photometrically calibrated. In Fig.~\ref{fig:color}, we show a
color composite image of the filters L$'$, Br$\alpha$ and
Pf$\gamma$.

\begin{figure*}
 \begin{center}
     \includegraphics[width=15.cm]{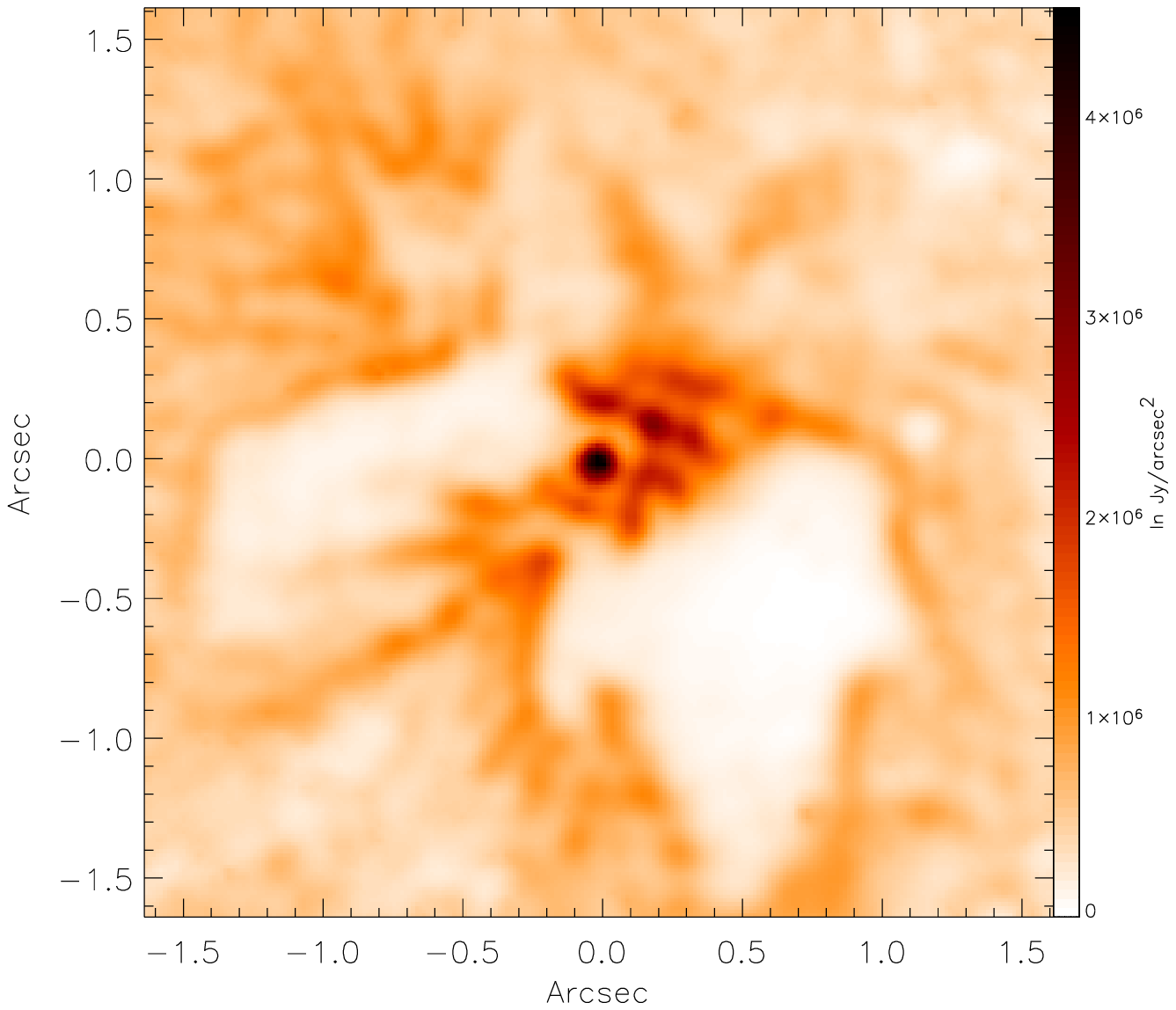}
\end{center}
  \caption[]{\label{fig:bracket} Pf$\gamma$ deconvolved image. The
  resolution achieved is in the order of 60~mas.
  To enhance the contrast, the image I$^{1/4}$ is shown. The color scale is expressed in
Jy/arcsec$^2$. Taken into account the large error bars of the
photometry, this scale is only indicative of the flux.
  }
\end{figure*}

The narrow-band images were deconvolved using the Richardson-Lucy
algorithm (1974) using as Point Spread Function (PSF) the star
HD~101104 observed immediately after the source acquisition. The
seeing during the 1h narrow-band images observations was stable,
typically 0.5 arcsec and the measured FWHM of the PSF at 3.74 and
4.05$\mu$m is 97 and 107~mas respectively, i.e. very close to the
diffraction limit of the telescope. By contrast, the FWHM of the
central object in Eta Car images in J, H and Ks is 65, 74 and
77~mas respectively, to be compared with the diffraction limit of
33, 43 and 57~mas respectively. We applied only 40 iterations to
enhance the spatial resolution and contrast of the images,
stopping before the appearance of any severe artifacts. The
resulting Pf$\gamma$ image is shown in Fig.~\ref{fig:bracket}. The
quality of the deconvolution process can be judged by the
comparison of the raw images and the deconvolved ones at
iterations 10 and 40 in Fig.~\ref{fig:NACOdeconv}.

The deconvolved images in the two filters are very similar apart
from the larger extension of the central object at 4.05~$\mu$m.
This is obviously an artefact of the deconvolution due to the fact
that in the Br~$\alpha$ filter, the 4-6 brightest pixels have
entered the non-linearity regime of the detector. Therefore the
central object differs from the true telescope PSF referenced with
the observation of HD~101104. The distortion of the central peak
mimics the flux emitted from a resolved object with the central
object appearing larger in the Br~$\alpha$ filter than in the
Pf$\gamma$ one (where the FWHM of the peak is about 60~mas, i.e.
60\% of the diffraction limit). This effect is localized and does
not affect the rest of the deconvolved image. Indeed one can check
in Fig.~\ref{fig:NACOdeconv} that all the structures are in common
between both filters.

We attempted to flux-calibrate the NB\_374 and NB\_405 images by
using the AO calibrator, HD~101104 observed immediately after Eta
Car. HD~101104 is a M4III star (Dumm \& Schild 1998) which has
been chosen for brightness considerations and not with the purpose
of photometric calibration. Hence this target is not well suited
for such a task but we attempted anyway to calibrate the flux
received by the narrow-band filters. From the typical intrinsic
color K-L$'$=0.21 of an M4III star, and measured K-band magnitude
of m$_{K}$=0.0$\pm$0.1., its L$'$ magnitude is estimated to be
m$_{L'}$=$-$0.2$\pm$0.1. Within the L' filter, the 3.74~$\mu$m
region of a M4 star is relatively free from lines, but the
4.05~$\mu$m region is strongly affected (Fluks et al. 1994).
Therefore, the intrinsic color of HD~101104 within the NB\_374
filter is about K-NB\_374=0.15$\pm$0.1, which means that the
magnitude of HD~101104 within this filter is
m$_{3.74}$=$-$0.15$\pm$0.1.

\begin{figure}
  \begin{center}
     \includegraphics[width=8.5cm]{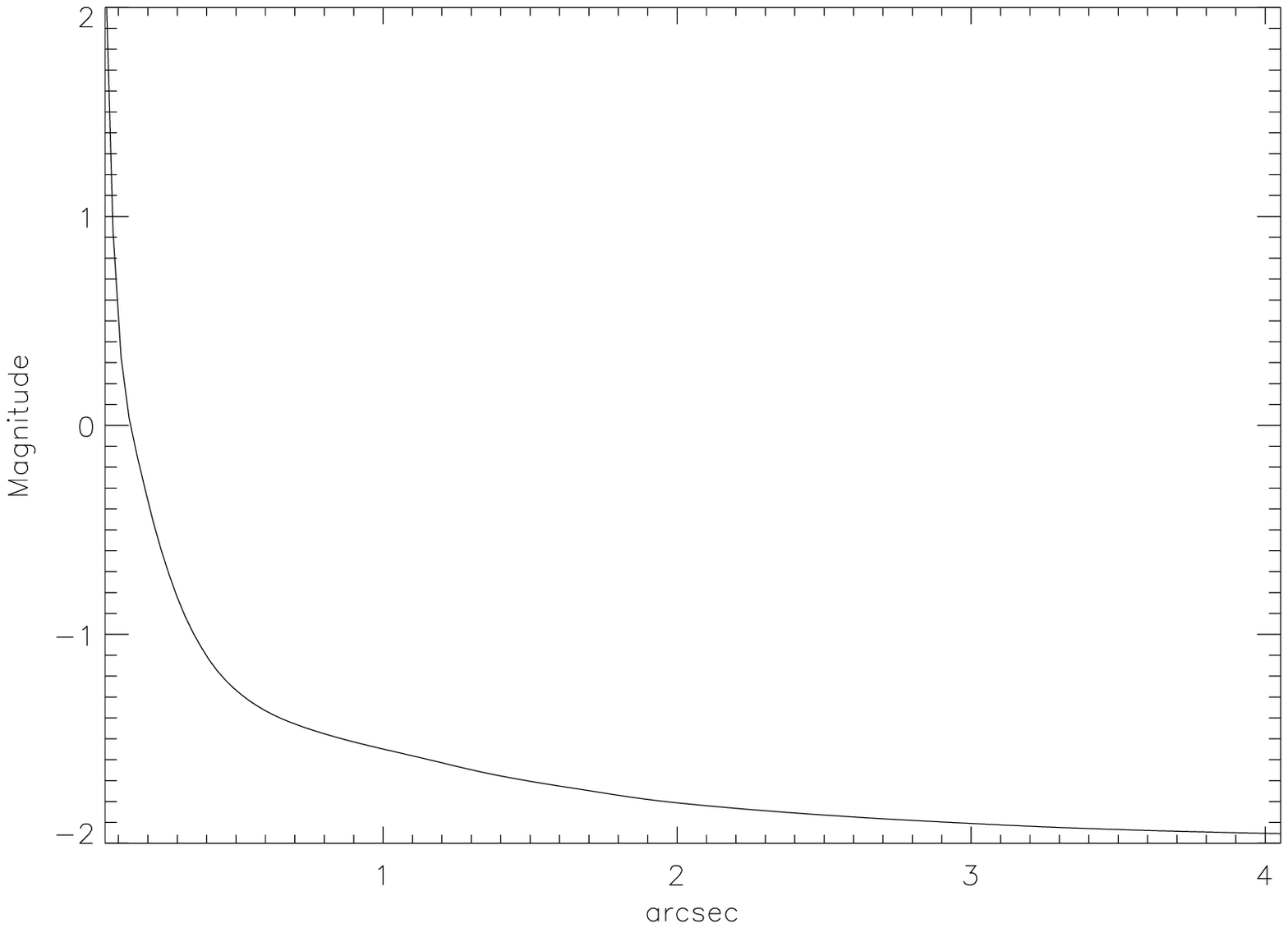}
   \end{center}
  \caption[]{\label{fig:NACOphot} Evolution of the L' magnitude
  based on aperture photometry with increasing radius. The
  integrated flux within a circle of 3$\arcsec$ is
  L$'$=-1.85$\pm$0.2.
   }
\end{figure}

\begin{figure*}
  \begin{center}
\includegraphics[width=16cm]{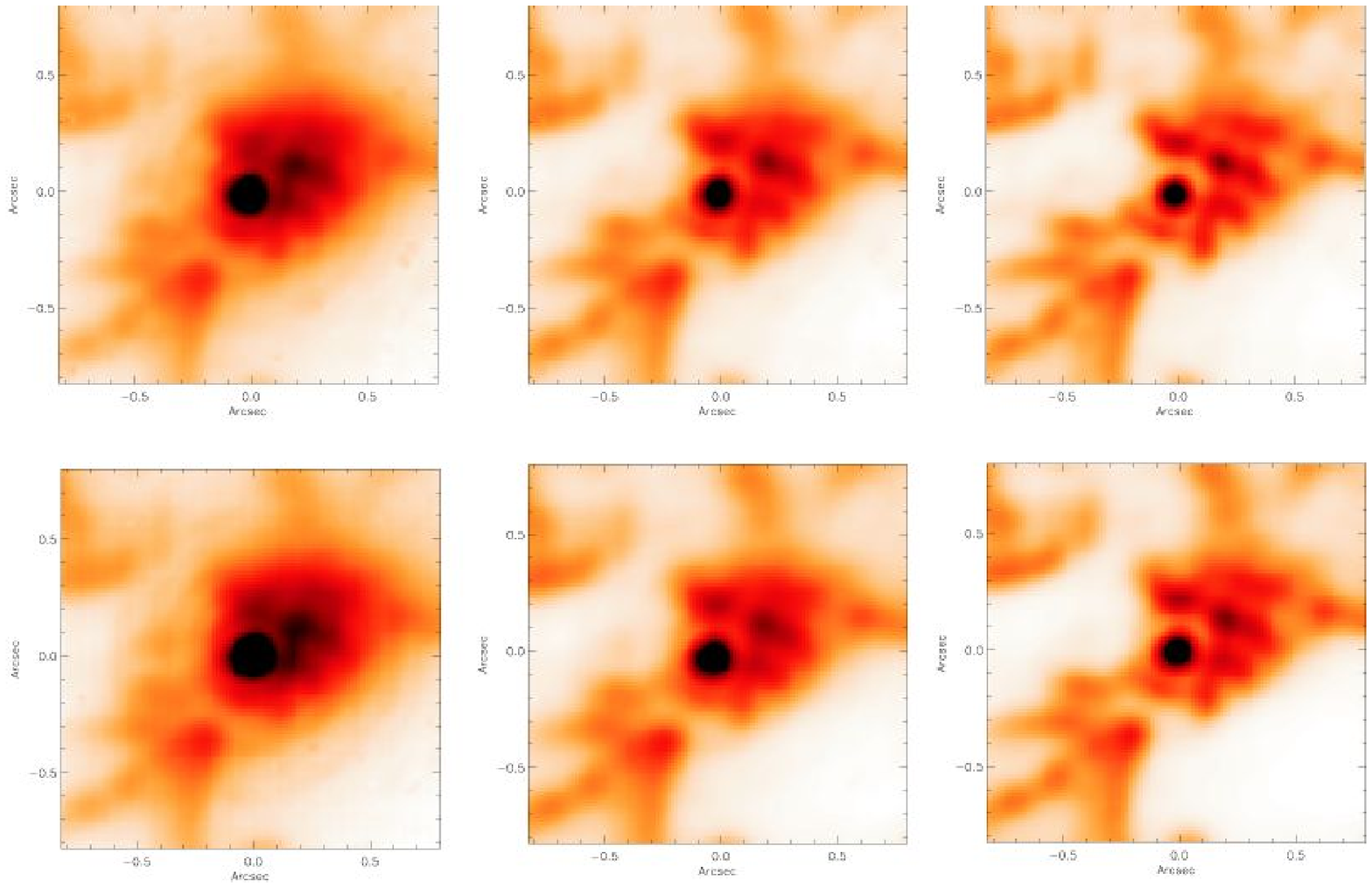}
   \end{center}
  \caption[]{\label{fig:NACOdeconv} Zoom into the deconvolved images from the NB\_3.74 (up) and NB\_4.05 (bottom) filters.
  The raw images are shown in the left side, the deconvolved
  images at iteration 10 and 40 are shown in the middle and the in the right side.
  The resulted images from the two filters are fairly similar except for the size of the central
  source which is 25\% larger in Br$\alpha$ due to a slight non-linearity of the
  detector at high flux regime.
   }
\end{figure*}

We estimated the L$'$ magnitude of Eta Car to be
L$'$=-1.85$\pm$0.2 within a circle of 3$\arcsec$, based on the
flux received in the L$'$ filters in the non-saturated regions and
also from the flux received in the 3.74~$\mu$m filter. The
evolution of the magnitude with the encircled flux is shown in
Fig.~\ref{fig:NACOphot}. A non-negligible flux is of course
emitted outside the studied regions but the dynamic reached with
the NACO short exposures is too limited to allow a good
photometric estimation outside this radius and the photometry
computed here is probably underestimated. The quasi-simultaneous L
band magnitudes (centered at 3.45~$\mu$m and not at 3.8~$\mu$m)
from Whitelock et al. 2004 are -1.737 for JD=2452603.60 (November
25, 2002) and -1.761 for JD=2452661.57 (January 22, 2003). By
scaling the PSF flux to the flux of the central source we estimate
the stellar contribution in the NB\_374 filter to be 520$\pm$70Jy
or m$_{3.74}$=$-$0.8$\pm$0.3.

\subsection{MIDI observations}
\label{sec:MIDIobs}
\subsubsection{Observing sequence}
\label{sec:obsseq} Eta Car was also observed with MIDI (Leinert,
Graser et al. 2003a and 2003b) the mid-infrared recombiner of the
Very Large Telescope (VLT). The VLTI/MIDI interferometer operates
like a classical Michelson interferometer to combine the MIR light
(N band, 7.5 - 14~$\mu$m) from two VLT Unit Telescopes (UTs). For
the observations presented here, the UT1 and the UT3 telescopes
were used, separated by 102~m with the baseline oriented
56$^\circ$ (E of N).

The observing sequence, typical of interferometric measurements,
is influenced by the design of the instrument (Leinert et al.
2003a, 2003b, Przygodda et al. 2003). The chopping mode (f=2Hz,
angle -90 degree) is used to visualize and accurately point at
the star.  The detector pixel size projected on the sky is
98~mas (measured by observations of close visual binaries) and the
field of view (FOV) is limited to 3$"$. The number of frames
recorded for each image was generally 2000, and the exposure time
per frame is by default 4~ms to avoid fast background saturation.
If the result of the centering is not satisfactory, the procedure
is started again.

\begin{table}[h]
 \caption{\label{tab:MIDI}Journal of observations with MIDI/UT1-UT3. The
 phase within the 5.52-year cycle
 is computed from the ephemeris of Daminelli et al. (2000).}
\vspace{0.3cm}
\begin{center}
 \begin{tabular}{llcc}
 \hline
  \hline
 Star & Template & Time &Frames \\
 \hline
\multicolumn{4}{c}{23/24-02-2003, JD=2452695, $\Phi=0.91$, B=75m, $\Theta=60^\circ$}\\
  \hline
  $\eta$ Car& Search. undispersed&T07:06:22&9000\\
  $\eta$ Car& Track. undispersed&T07:09:25&9000\\
 $\eta$ Car& Track. dispersed&T07:15:01&9000\\
 $\eta$ Car& Search. undispersed&T07:20:01&2800\\
  $\eta$ Car& Search. undispersed&T07:24:03&2800\\
 $\eta$ Car& Phot. UT1 disp.&T07:36:42&400\\
 $\eta$ Car& Phot. UT3 disp.&T07:38:59&400\\
\hline
\multicolumn{4}{c}{12/13-06-2003, JD=2452893, $\Phi=0.96$, B=74m, $\Theta=62^\circ$}\\
 \hline
 $\eta$ Car& Track. dispersed&T00:31:24&9000\\
 $\eta$ Car& Track. dispersed&T00:35:43&10000\\
 $\eta$ Car& Track. dispersed&T00:39:17&7500\\
 $\eta$ Car& Phot. UT1 disp.&T00:43:53&3000\\
 $\eta$ Car& Phot. UT3 disp.&T00:45:55&3000\\

 \hline

 \hline
\multicolumn{4}{c}{14/15-06-2003, JD=2452806, $\Phi=0.96$, B=78m, $\Theta=56^\circ$}\\
 \hline
HD~120323& Acquisition N8.7$\mu$m&T23:08:03&2000\\
HD~120323& Acquisition N8.7$\mu$m&T23:09:16&2000\\
HD~120323& Track. dispersed&T23:18:08&12000\\
HD~120323& Phot. UT1 disp.&T23:23:24&3000\\
HD~120323& Phot. UT3 disp.&T23:25:28&3000\\
$\eta$ Car& Acquisition N8.7$\mu$m&T23:47:45&2000\\
$\eta$ Car& Acquisition N8.7$\mu$m&T23:48:51&2000\\
 $\eta$ Car& Track. dispersed&T23:55:47&12000\\
 $\eta$ Car& Phot. UT1 disp.&T00:02:10&3000\\
 $\eta$ Car& Phot. UT3 disp.&T00:03:49&3000\\
HD~148478&Acquisition N8.7$\mu$m&T00:14:53&2000\\
HD~148478&Acquisition N8.7$\mu$m&T00:17:01&2000\\
HD~151680&Acquisition N8.7$\mu$m&T00:24:12&2000\\
HD~151680&Acquisition N8.7$\mu$m&T00:25:23&2000\\
HD~151680& Track. dispersed&T00:32:49&12000\\
HD~151680& Phot. UT1 disp.&T00:38:22&2000\\
HD~151680& Phot. UT3 disp.&T00:40:08&2000\\
 HD~167618&Acquisition N8.7$\mu$m&T01:18:33&2000\\
 HD~167618&Acquisition N8.7$\mu$m&T01:19:41&2000\\
 HD~167618& Track. dispersed&T01:38:28&12000\\
 HD~167618& Phot. UT1 disp.&T01:33:15&3000\\
HD~167618& Phot. UT3 disp.&T01:35:28&3000\\
HD~168454&Acquisition N8.7$\mu$m&T02:35:55&2000\\
HD~168454&Acquisition N8.7$\mu$m&T02:37:12&2000\\
HD~168454& Track. dispersed&T02:40:47&12000\\
HD~168454& Phot. UT1 disp.&T02:50:18&3000\\
HD~168454& Phot. UT3 disp.&T02:52:34&3000\\

\hline
  \end{tabular}
\end{center}
 \end{table}

Then, the MIDI beam combiner, the wide slit (0$\farcs6
\times$3$\arcsec$), and the NaCl prism are inserted to disperse
the light and search for the fringes by moving the VLTI delay
lines. The resulting spectra have a resolution $\lambda$/$\Delta
\lambda \sim$30. When searching for the fringe signal, the large
delay line of the VLTI is moved to compensate for Earth rotation
and atmospheric delays, while the MIDI internal
piezo-driven delay line is driven in scans to create
the fringe pattern. Once the fringes are found a file is recorded
while MIDI is self-tracking them. Finally, two other files are
recorded for the photometry. In the first file, one shutter only
is opened, corresponding to the calibration of the flux from UT1
and the flux is then divided by the MIDI beam splitter and falls
on two different regions of the detector. The total flux is
determined separately by chopping between the object and an empty
region of the sky, and determining the source flux by subtraction.
Then the same procedure is applied with UT3.
\begin{figure*}
  \begin{center}
    \includegraphics[height=6.8cm]{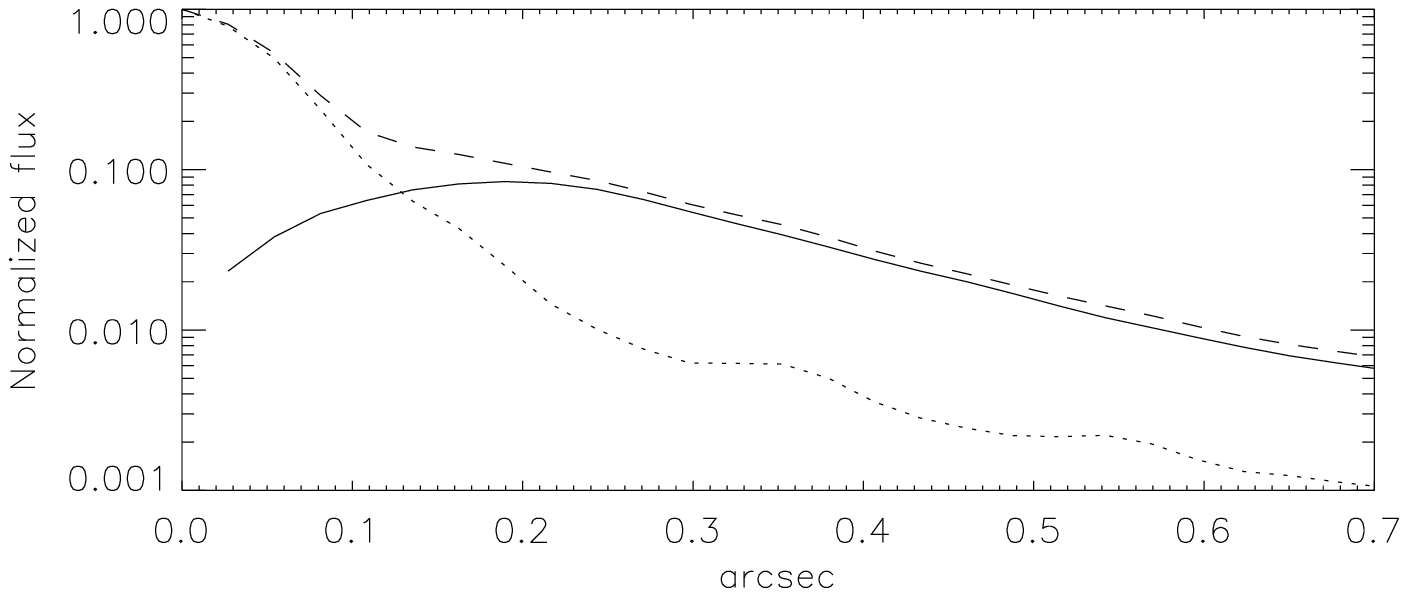}
\includegraphics[height=6.8cm]{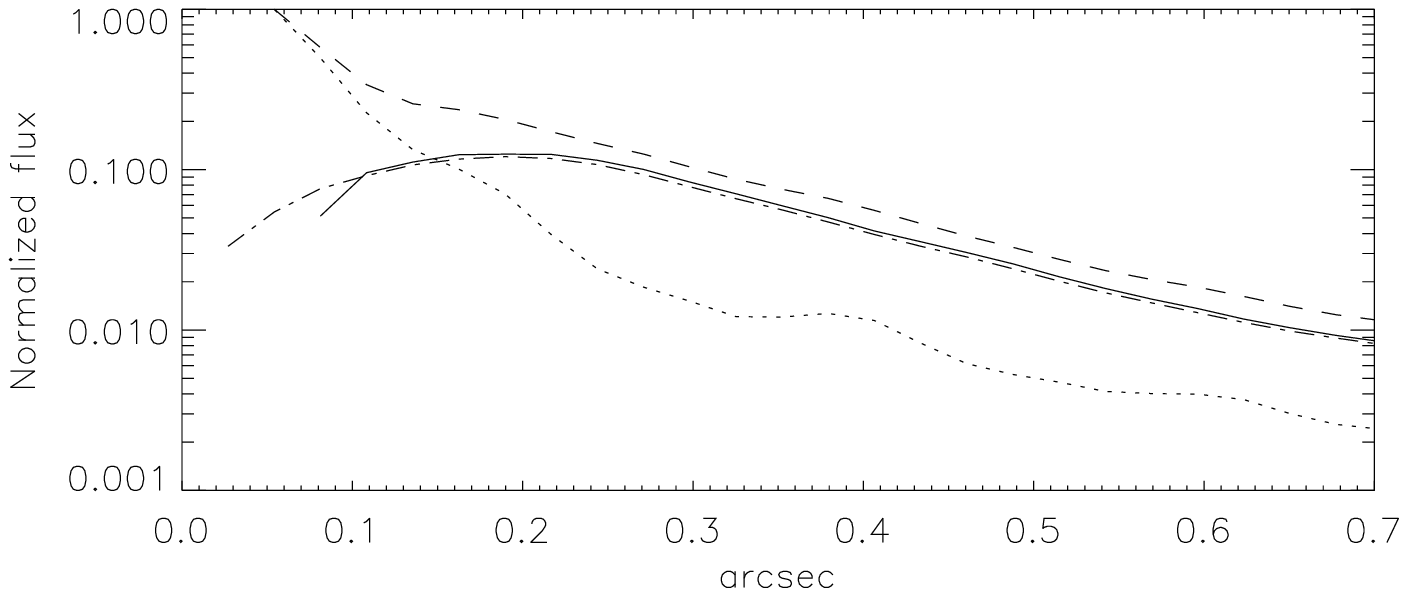}
  \end{center}
  \caption[]{\label{fig:radial}On the top, radial flux normalized to the peak with NB\_3.74
  filter for Eta Car (dashed line),
  and the PSF HD~101104 (dotted line) and their subtraction (solid line).
  Close to the source, a strong decrease of the flux is clearly seen that we attribute to
the dust sublimation region at about 0$\farcs$1 (230 AU). On the
bottom panel the same treatment is applied to the NB\_4.05 images.
However, the flux reference is chosen to be at $\sim$0$\farcs$05
(2 pixels) from the peak in order to account for the non-linearity
of the detector at maximum. The dashed-dotted curve is the
residual of the NB\_3.74 filter subtraction scaled for comparison.
The two curves agree fairly well and the contribution of the
Br$\alpha$ line is hardly visible.}
\end{figure*}

A list of all observations is presented in Table~2. The scientific
observations have been mainly conducted in the night of the 16th
and 17th of June 2003. Unfortunately, Eta Car was only observable
at the very beginning of the night and has consequently been
observed with two close sky projected baselines of 74~m and 78~m
and PA=57$^\circ$ and 62$^\circ$ respectively. We also report some
observations which were performed in February 2003 during the
first MIDI commissioning run (see Sec.~\ref{sec:spatfringes}).
These observations were carried out when the MIDI fringe tracker
was not performing well and the sensitivity is quite limited
compared to the measurements performed in June. However, some data
recorded in undispersed mode are of particular interest and
are presented here (Sect.~\ref{sec:spatfringes}).

\subsubsection{Acquisition images}
\label{sec:acqim}  Custom software, written in the IDL language
was used to reduce the images, spectra and fringe data. MIDI is a
relatively new and unique instrument. The MIDI data reduction is
described more extensively in Leinert et al. (2004) and a devoted
paper is in preparation.

The first step of the reduction is to read in the acquisition
datasets, average the frames on the target and the frames on the
sky, and subtract the average sky frame from the average target
frame. Despite the high number of optical elements in the
VLTI/MIDI system (33 in total), the quality of the 8.7~$\mu$m
images is comparable to the best Mid-IR images published to date
(Smith et al. 2002a). The spatial resolution has been slightly
increased by performing a deconvolution using 40 iterations of the
Lucy-Richardson algorithm and the result is shown in
Fig.~\ref{fig:im87}. The spatial resolution reached after the
treatment is about 150~mas.
Most MIDI targets are  unresolved by a
single 8~m telescope; thus many PSF samples are available for testing the
quality of the deconvolution process. We used as PSF reference the
acquisition images of HD~120323 (2 Cen, M4.5III, m$_{8.7}$=-1.8),
HD~148478 ($\alpha$~Sco, M1.5Ib, m$_{8.7}$=-4.34) and HD~151680
($\epsilon$~Sco, K2.5III, m$_{8.7}$=-0.37 extrapolated).

The 8.7~$\mu$m magnitudes have been extracted from the Catalog of
Infrared Observations (Edition 5, Gezari+ 1999). Accurate 8.7~$\mu$m
photometry of the MIDI acquisition images is  difficult.
First, the acquisitions from Eta Car and $\alpha$~Sco were
slightly saturated, which affects the linearity of the detector
response but also its local offset. Then, due to a pupil mismatch,
the FOV and also the background level are different between the
two telescopes. For UT3, the FOV was about 3$\arcsec
\times$2$\arcsec$ while for UT1 the FOV was less than 2$\arcsec
\times$2$\arcsec$. We end up with an integrated 8.7~$\mu$m
magnitude within UT3 and UT1 FOVs of -5.65$\pm$0.3 and
-5.45$\pm$0.3 respectively. One of the MIDI acquisition images
from UT3 is shown in Fig.~\ref{fig:slit} and the best deconvolved
8.7~$\mu$m image is shown in Fig.~\ref{fig:DeconvMIDI}. The flux
scale given in Fig.~\ref{fig:im87} should be considered only as an
indication due to the large errors mentioned above and also due to
the further difficulties in the deconvolution process.

\subsubsection{Dispersed photometry}
\label{sec:dispphot} The second reduction step consists of reading
the dispersed photometric datasets used for the calibration of the
contrast of the dispersed fringes. We use the same procedure to
average the frames on the target and the frames on the sky, and then
subtract the average sky frame from the average target frame. Eta
Car is a complex extended object which requires a dedicated
procedure of reduction. In the following we describe the
'classical' data reduction of MIDI data applied for sources
unresolved by a single dish telescope, i.e. for the calibrators.

In the averaged, subtracted frame, the wavelength axis is oriented along
the horizontal detector axis. For each detector column, the
vertical centroid and width of the spectrum are estimated by
fitting a Gaussian function to the peak. The centroid position in
all illuminated columns is fitted with a quadratic polynomial as
a function of column number, while the width is fitted by a linear function.
This procedure is carried
out on both photometric datasets (corresponding to telescope UT1
and UT3 respectively). Both fits are averaged and used to create a
2-dimensional weighting mask consisting of a gaussian
function with the average position and width of the spectra for each column.
This mask is applied to both photometric and interferometric data to
supress noise in regions where few source photons fall.
(Sect.~\ref{sec:dispcor}).

We used HD~167618 ($\eta$ Sgr) and HD~168454 ($\delta$ Sgr) as
calibrators for the dispersed photometry. They are secondary flux
calibrators which were observed several times during the June run
and have been calibrated using two primary calibrators ($\alpha$
Aql and HD~177716) for which very good quality spectra are
available (Cohen et al. 1998). The airmass correction is extracted
from the observations of HD~168454 at two different airmasses.

Considering the very good atmospheric conditions encountered
during this observing run, the N-band images are
close to the diffraction limit. This implies that we can obtain
about 8-10 independent spectra of the nebula with a mean spatial FWHM of
about 250~mas. This information is valuable for the study of the
nebula dust content and provide complementary information to the
correlated flux study described in Sec.~\ref{sec:correlated}.

A set of spatially resolved spectra of Eta Car were extracted
along the slit direction using a
modified version of the MIDI photometric algorithm. A `point source'
weighting mask was constructed from a calibrator, and then
shifted along the slit to extract a sequence of spectra from
Eta Car itself.  The stability of MIDI
within its cryostat permits such a procedure. The purpose is to
avoid a bias on the trace due to the shift of the emission
photocenter with wavelength in the N band towards the north-west,
due to the increasing contribution of colder dust situated in the
Weigelt complex (Smith et al. 2002a).

The slit was positioned at PA=140$^\circ \pm$3$^\circ$, i.e. very
close to the nebular principal orientation. Nine spectra with
sufficient SNR have been extracted separated by 400 mas, the five
central ones using a gaussian weighting function with a FWHM of
200 mas and the four external ones with a slightly larger beam in
order to increase the SNR (which implies a slight cross-talk
between the beams). The parameters of the apertures are reported
in Tab~\ref{tab:dust} and the spectra are presented in
Fig.~\ref{fig:specmodel}.

\subsubsection{Undispersed correlated flux}
\label{sec:undispcor} The spatial distribution of the fringes
detected by MIDI with the 8.7~$\mu$m filter is shown in
Fig.~\ref{fig:spatfringes}. The data were recorded during February
commissioning time when the main observing modes of MIDI had not
yet `crystallized'. This figure is very interesting because it is
one of the first wide field interferometric detections of fringes
reported to date. The fringes are detected by measuring the
fluctuation of the detector power, pixel-by-pixel, as the OPD is
scanned. The level of the detector, background and sky noise are
clearly identified and well constrained by the multiple tests
performed on the sky during MIDI commissioning. By choosing the
frames for which fringes are detected we can localize the fringe
signal to an area of about a PSF size. Since fringes are detected
only in the common part of the beams coming from the individual
UTs, the spot visible in calibrators exhibits a sightly smaller
FWHM than in the non-interferometric acquisition images. We have
checked that the calibrators observed before and after Eta Car
have a similar extension by performing a 2D gaussian fit. Due to
the individual pointing errors from each telescope, the overlap
spot is not necessarily perfectly symmetric, but no asymmetry
larger than 15\% is observable. In contrast to calibrators, the
fringes from Eta Car are more extended than the acquisition PSF.
To verify the extension of this signal we have first removed a
noise pattern extracted at OPD positions farther than 1 millimeter
away from the white light fringe (OPD=0), i.e. at a location where
no fringes are present. A typical spot extracted from a calibrator
was then centered on the maximum, scaled and subtracted from the
figure. This technique is similar to the one used to remove a PSF
from images.

\subsubsection{Dispersed correlated flux}
\label{sec:dispcor} Each frame of the fringe data is reduced to a
one-dimensional spectrum by multiplying it with the mask,
integrating in the direction perpendicular to the spectral
dispersion, and finally subtracting the two oppositely phased
output channels of the beam combiner from each other.
The spectra from each scan with the
piezo-mounted mirrors were collected and fourier-transformed with
respect to OPD. The
fringe amplitude at each wavelength is then obtained from the
power spectrum. We typically summed four pixels in the dispersion
direction to improve the signal-to-noise ratio.

The correlated flux varies from scan to scan due to the
variable overlap of the two telescope beams.  The scans used to
estimate the  flux were selected based
on the white-light fringe amplitude, i.e.\ the fringe amplitude
that is seen after integrating over all usable wavelengths. The
histogram of all white-light fringe amplitudes within a fringe
track dataset usually shows a small peak near zero, and a broad
peak at higher amplitudes. We interactively set a threshold just
below this broad peak, and average the fringe amplitudes of all
scans with a white-light fringe amplitude higher than this
threshold to give the final fringe amplitude as a function of
wavelength. The raw visibility is obtained by dividing the fringe
amplitude by the total photometric flux. The calibrated visibility is obtained
by dividing the raw visibility of an object by that of an
unresolved calibrator. The photometrically calibrated flux
creating the fringes is called correlated flux.

The MIDI reduction software has been modified to allow it to
handle spatially extended fringes in the slit direction. The
extraction mask, which is usually wide in the slit direction to
include all the light from the sources has been narrowed to cover
no more than 3-4 pixels (i.e. 0$\farcs$3-0$\farcs$4) along the
slit. We used the set of masks created for the photometric study
described in table~\ref{tab:dust}. At first we checked that
the calibrated visibilities of calibrators derived with this mask
were identical to the ones derived with the normal width. It
turned out that the instrumental visibility is slightly higher
(about 5-10\%, depending on the wavelength) when the mask is
narrower. This bias is perfectly corrected when the visibilities
are calibrated, i.e. when the raw visibility of the science object
is divided by the calibrator visibility.

A binning of 6 pixels in the dispersion direction was used to
increase the signal, providing a spectral resolution of about 15.
The visibilities were calibrated and multiplied by the flux
calibrated spectra shown in Fig.~\ref{fig:specmodel}. The aperture
5 is placed at the maximum of correlated flux which coincides with
the peak of the deconvolved 8.7~$\mu$m image.

\section{Description of the images}
 \label{sec:images}
Our NACO observations cover at unprecedented spatial resolution
the spectral region of transition between the optical and near-IR,
dominated by scattering processes while the MIDI observations
cover the mid-IR thermal emission from the dust regions. The
general appearance of the J, H, Ks images is indeed very different
from that of the L$'$ and the 8.7~$\mu$m images.

In the near-infrared, the central 1$\farcs$5 of the nebula are
dominated by a point source. A complex 'butterfly' shape
morphology in the immediate vicinity emerges clearly only at
around 2~$\mu$m, though some features can already be traced
at shorter wavelengths (Fig.~\ref{fig:continuum}).

%The recent mid-IR observations of Smith et al. (2002a, 2003a)
%showed an unexpected lack of azimuthal symmetry and they conclude
%that if a torus was preexisting the outburst, it is now at least
%severely disrupted.

\begin{figure}
  \begin{center}
     \includegraphics[height=7.8cm]{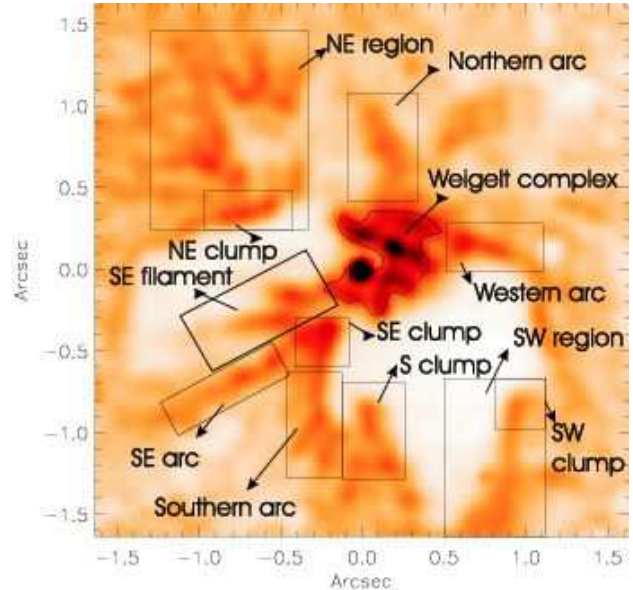}
   \end{center}
  \caption[]{\label{fig:sketch} Location of interesting regions in
  the 'butterfly' dusty nebula close to Eta Car. The naming convention is partly based on
  Fig. 4 of Smith et al. (2003a).
   }
\end{figure}

In Fig.~\ref{fig:sketch}, we have labelled the structures seen in
our deconvolved NACO 3.74~$\mu$m image in order to guide the
discussion on the geometry of the dusty nebula.

\subsection{Weigelt blobs and other dusty clumps}
The core (central 1$\arcsec \times$ 1$\arcsec$ aperture) is
dominated in the northwest by a region of large dust content which
was not resolved by Smith et al. (2003a) but is clearly
visible in our NACO data which we call the 'Weigelt' complex
(Weigelt \& Ebersberger 1986; Hofmann \& Weigelt 1988). The
'Weigelt complex' dominates the stellar flux in the mid-infrared,
shifting the core photocenter from the star to about 0$\farcs$3
north, as seen by the difference between the raw and deconvolved
MIDI images at 8.7 $\mu$m (Fig.~\ref{fig:slit} and
Fig.~\ref{fig:DeconvMIDI}). The Weigelt blobs are small, dense
knots of dust and gas within about 1000 AU, i.e 0$\farcs$3
northwest from the star.

\begin{figure*}
  \begin{center}
\includegraphics[width=8.7cm]{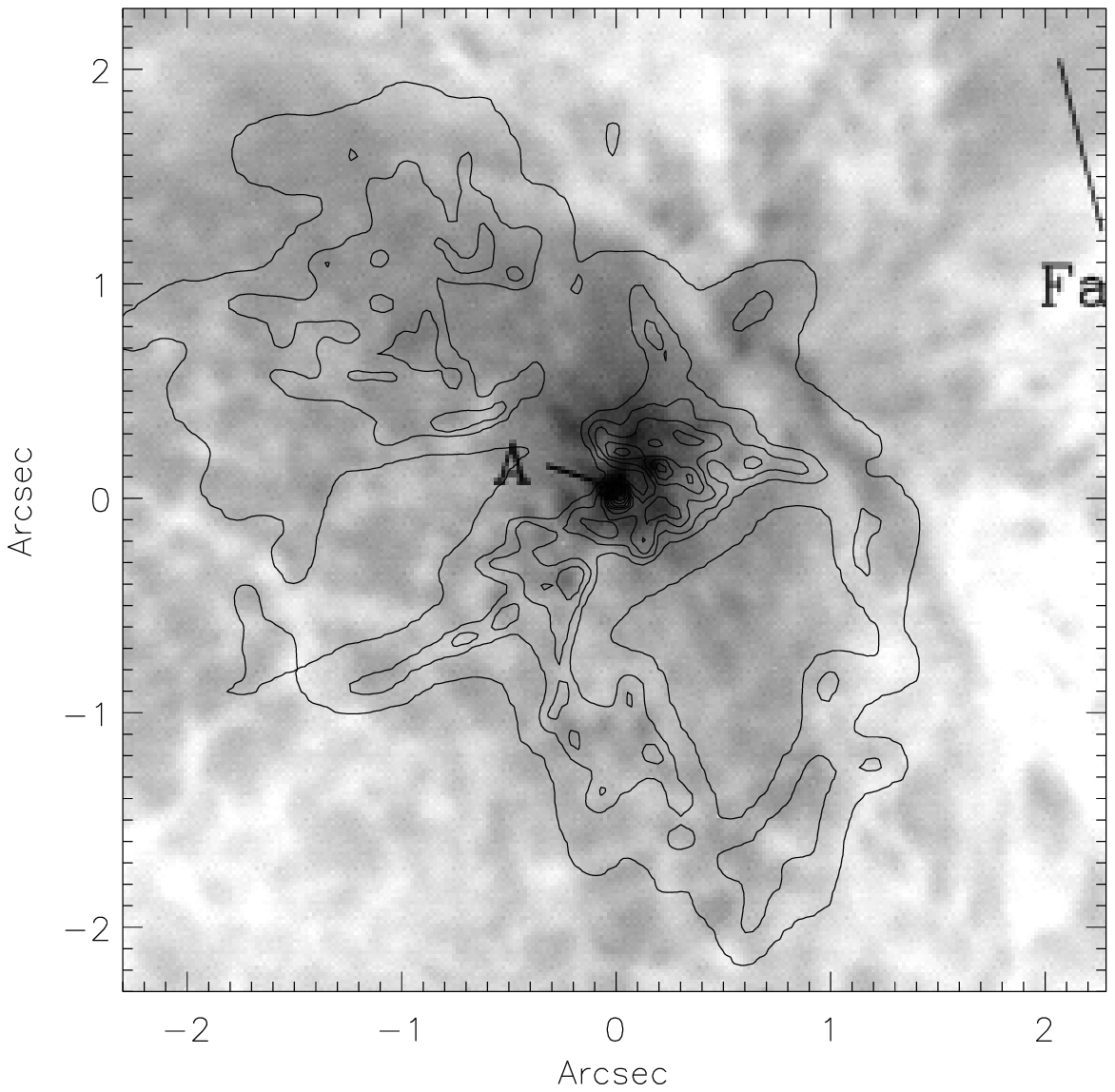}
    \includegraphics[width=8.7cm]{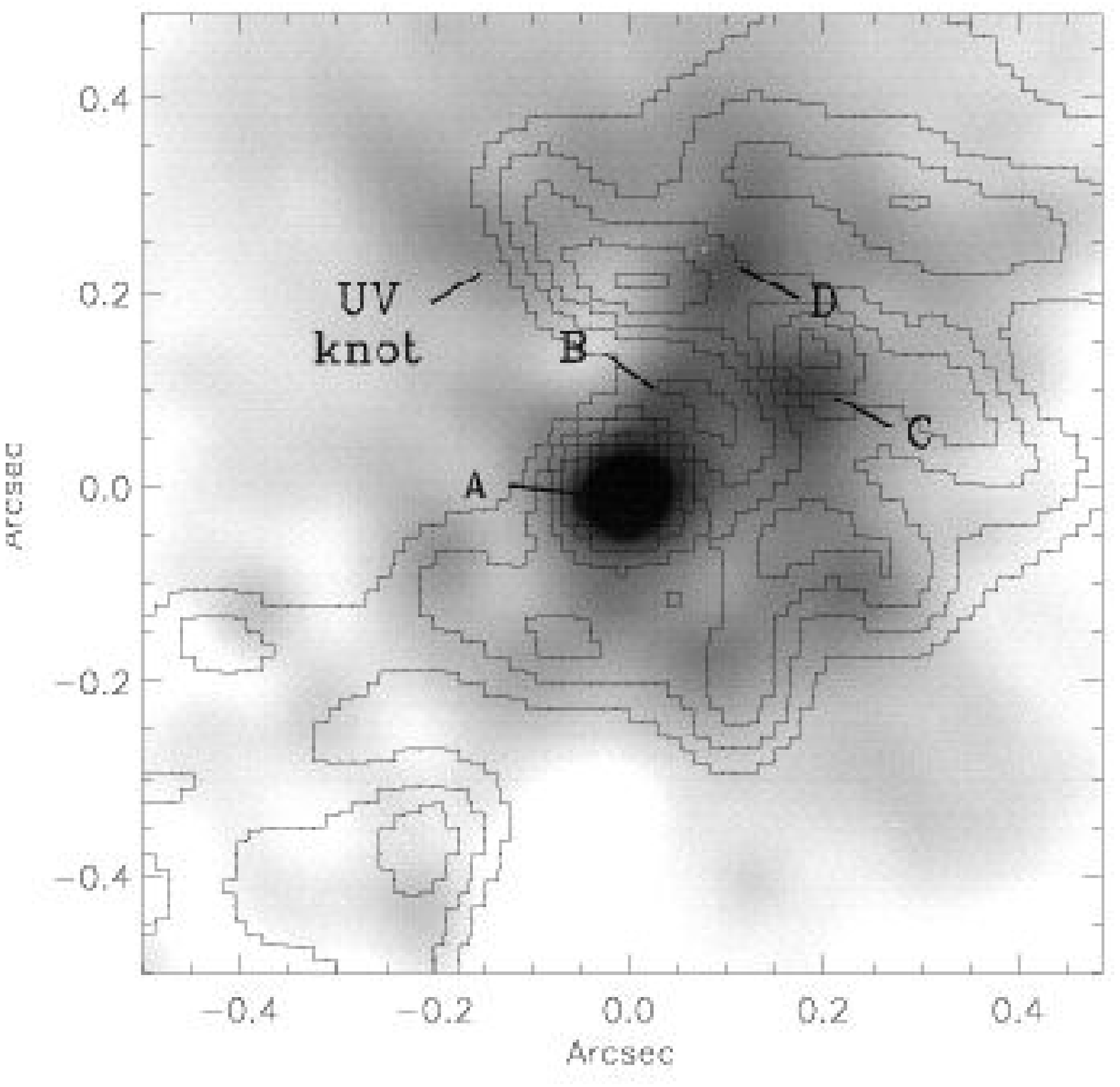}

  \end{center}
  \caption[]{\label{fig:morse}Comparison of the 3.74 $\mu$m deconvolved
  NACO image (in contours) with the highest resolution HST images (background, from Morse et al. 1999)
  at large (left) and small (right) scales. The location of the
  dusty clumps does not coincide with that of the Weigelt blobs, but their structure is somehow
  complementary to the image. The optical/UV blobs are probably hot regions less shielded from
  the central star's UV flux which could coincide with the dust clump surfaces facing the star. This
  is particularly true for Weigelt blobs C and D.}
\end{figure*}

The positions and hence the velocities of the Weigelt blobs were
studied by further speckle observations about 10 years after the
initial ones (Weigelt et al. 1995, 1996) and also by HST imaging
and spectroscopy (Davidson et al. 1997, Dorland et al. 2004, Smith
et al. 2004a). These observations demonstrated that the ejecta are
moving slowly (less than 50 km.s$^{-1}$ from the star) on the
equatorial plane. This means that our image of the Weigelt complex
seen at 4~$\mu$m can probably be compared to the optical ones
detected some years ago. An attempt of this kind is shown in
Fig.~\ref{fig:morse} by using the images published in Morse et al.
1998.

The position of the largest structures (clumps) can be measured
accurately from our images. The brighter clump can be related to
the Weigelt clump C, but the clump B is clearly absent. The
location of clump D is close but not coincident to a bright clump
in the `hook' shaped region directly north of the star. The hook
is also very close to a structure called the UV knot in the image
published by Morse et al. 1998 (Fig.~\ref{fig:morse}). It must be
stressed out that this comparison is based on images separated in
time. However Dorland et al. (2004) have confirmed that the proper
motion of these structures is at most $\simeq$5~mas per year. In a
span of 5 years the clumps have moved by at most 25~mas, i.e. less
than one pixel on the NACO detector. We are convinced that the
structures seen in the NIR are correlated to the ones seen in the
visible or UV. The visible structures, dominated by scattering,
trace the  {\it walls} of the dense clumps of dust. We propose to
call the two brightest NIR clumps to the north-east of the star
clumps C' and D' since they do not coincide with the visible ones
but are probably related to them.

\begin{table*}
 \caption{\label{tab:clumps}Separation and position angle measurements with respect to the star
 of several blobs seen in our images.}
\vspace{0.3cm}
\begin{center}
 \begin{tabular}{lcccccc}
\hline \hline

 &\multicolumn{2}{c}{Blob C'/C}&\multicolumn{2}{c}{Blob D'/D}&\multicolumn{2}{c}{Blob SE}\\
&Separation & P.A. & Separation & P.A. &Separation & P.A. \\
Image & (mas) & (deg) & (mas) & (deg) &(mas) & (deg) \\
\hline
Ks&-&-&-&-&379$\pm$4&120.0$\pm$2\\
NB\_374& 233$\pm$13 & 300$\pm$15 & 218$\pm$14 & 352$\pm$17&418$\pm$13&121.9$\pm$3\\
NB\_405&237$\pm$9&298$\pm$11&216$\pm$9&359$\pm$14&401$\pm$21 & 123.4$\pm$3\\

%L$'$&-&-&-&-&&\\
8.7$\mu$m&-&-&-&-&603$\pm$38&136$\pm$12\\

%\hline
% &\multicolumn{2}{c}{Blob C}&\multicolumn{2}{c}{Blob D}&\multicolumn{2}{c}{Blob SE}\\
\hline
F550M$^1$ & 224$\pm$3 & 299$\pm$3 & 259$\pm$3 & 335$\pm$3 & - & - \\
\hline

\multicolumn{7}{l}{{\small $^1$- Optical data from HST taken in
2002.786 from Smith et al. 2004a.} }

  \end{tabular}
\end{center}
 \end{table*}

Astrometric measurements of the Weigelt blobs C and D have been
recently reported by Dorland, Currie and Hajian (2004) and by
Smith et al. (2004a). Both used HST data to measure the relative
proper motions of blob C and blob D with respect to the star. The
position angles are consistent with linear, radial, ballistic
motions and no evidence of azimuthal motion was detected. The
weighted position angle measurements from Dorland et al. for blobs
C and D are PA$_C$ = 300.6$\pm$0.6 and PA$_D$ = 336.8$\pm$0.4
degrees. Dorland et al. used a least-squares three-parameter fit of
a two-dimensional Gaussian with fixed width and removed the strong
diffusive background by using a median filter method.

We applied a similar method on our NACO images, i.e. on the two
deconvolved images of the narrow band filters at 3.74~$\mu$m and
4.05~$\mu$m and on the Ks image (without median filtering). This
method was not possible for the saturated L$'$ filter. We also used
a two-dimensional Gaussian for the fits with the FWHM taken as a
free parameter. These measurements are a starting
point to a further monitoring of these structures by NACO. We
decided to concentrate on the two brightest blobs D' and C' and on
the bright southern clump that we call SE (Fig.~\ref{fig:sketch}).
The results are presented in table~\ref{tab:clumps}.

While Dorland et al. and Smith et al. propose different ejection
dates for the clumps of 1941 and 1890, respectively, their
measured values of 1934$\pm$20 and 1907$\pm$12 agree within
statistical errors. In visible and UV spectral regions, the blob
emission is dominated by the ablated halo while the dusty clumps
are traced by NACO. The consequence is a large variability in
shapes and hence centroid positions of the blobs seen with
different filters. NACO has the great advantage of providing the
location of the dust clumps in the NIR with a spatial resolution
comparable to that of the HST. The contribution of the scattered
light in L$'$ is drastically reduced compared to shorter
wavelengths. NACO measurements not yet able to provide
further constraints of the outburst but soon will be. More effort
should be put in  to decreasing the error bars of single position
measurements by using different methods of position determination,
as shown by Smith et al.. We thus advocate a monitoring of the
Weigelt complex by NACO at least during a span of 6 years which
would correspond also to the course of the 5.52-year motion of the
binary.

\subsection{Arcs and filaments}
\begin{figure}
  \begin{center}
\includegraphics[height=9.cm]{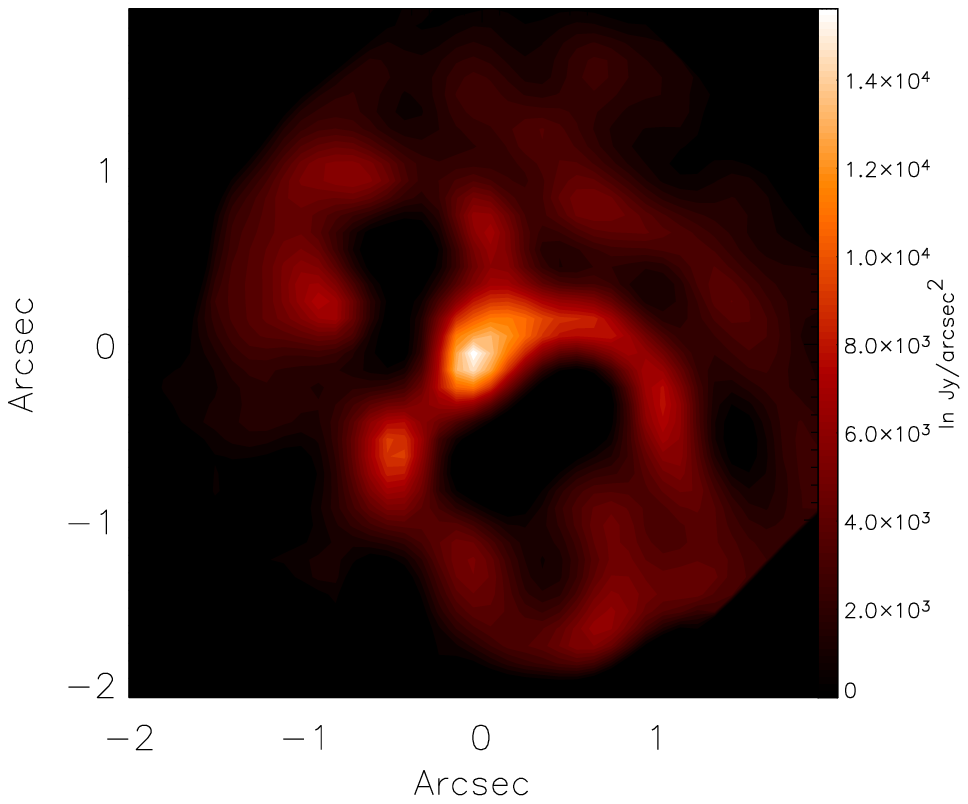}
  \end{center}
  \caption[]{\label{fig:DeconvMIDI}Deconvolved acquisition image of Eta Car with the 8.7~$\mu$m
filter (UT3). In order to enhance the contrast, the image
I$^{1/4}$ is shown but the color scale is expressed in
Jy/arcsec$^2$. The bright spot which emerges in this image is in
the position of the central star and at the location where a
strong correlated flux has been detected by MIDI (see
Fig.~\ref{fig:spatfringes}). The image has been de-rotated with
respect to Fig.~\ref{fig:slit} so that the north is up and east is
left.
  \label{fig:im87}}
\end{figure}
The geometrical aspect of the dusty nebula is impressive. It is
characterized in both the NACO and MIDI images by two particularly
dark regions in the east and south-west, and a third one in the
north-east where a faint nebulosity is visible, suggesting that
these regions are also relatively devoid of dust.

The resemblance between the MIDI deconvolved image
(Fig.~\ref{fig:DeconvMIDI}) and the NACO one is striking. A large
part of the regions denoted in Fig.~\ref{fig:sketch} can be
recognized (some of them interrupted by the limits of the MIDI
FOV): the Weigelt complex, the NE and SW regions, the Northern and
Western arcs, the S clump.

It must be pointed out that the brighter clumps discussed in the
previous section are just the emerged part of a fainter nebulosity
contained within well-defined borders of about
0$\farcs$5$\times$0$\farcs$5 shown in Fig.~\ref{fig:sketch}. This
triangle-like nebulosity seems to be connected to the south with a
fainter structure, the `SE filament' apparently aligned in the
same direction as the Weigelt blobs complex
(PA$\sim$300-330$^\circ$). There is no clear separation between
the bright northern Weigelt blobs complex and this SE filament.
Moreover, the bright Weigelt complex clearly embeds the star
itself and the most probable explanation of the faintness of the
SE filament is the small amount of material involved in this
structure. The Weigelt complex appears interrupted in the
north-east of the star giving birth to a `hook' region directly to
the North, reminiscent of the one detected in UV by Morse et al.
1998 (see also Fig. 7).

Of particular interest is the bright spot at about
0$\farcs$5-0$\farcs$8 southeast of the star, seen particularly well
in our 8.7~$\mu$m (also see Smith et al. 2003a) and in the L$'$,
Pf$\gamma$ and Br$\alpha$ images. This blob connects two
well-defined arcs: the Southern arc and the SE arc. The SE arc,
brighter in the NIR has already been denoted as `jet' from images
at lower resolution (see Rigault \& Gejring 1995 or Fig.1 in Smith
\& Gehrz 2000 for instance).

The Southern and SE arcs seem to partly hide the SE filament and
seem to be in front of it. Moreover they are connected in a
complex but traceable way to the northern arcs. These arcs are
apparently hidden (or embedded) in the north by the Weigelt
complex. This is particularly visible in the 3.74~$\mu$m and
4.05~$\mu$m images (Fig.~\ref{fig:NACOdeconv}), but some hints can
also be extracted from the MIDI image.

\subsection{Dust sublimation radius}
Our NACO Pf$\gamma$ and Br$\alpha$ deconvolved images show no
evidence for significant emission in the inner regions (see
Fig.~\ref{fig:radial}). The radius where the flux inflexion point
is located is about 130-170~mas in both deconvolved images. The
width of the empty regions is not symmetric around the source, it
is more extended to the north than to the south. It could be
argued that this gap is an artefact of the deconvolution process
but a decrease of emissivity is already visible in the radial
profiles shown in Fig.~\ref{fig:radial}, extracted from the raw
images. Moreover, we carried out further tests to verify the
reality of the feature. While increasing the number of iterations
by steps, we checked that at each time the gap remained stable
in size and shape. Until the appearance of strong artifacts
affecting the whole image, this feature behaves like any other:
its shape is slowly distorted, but its position remains essentially
unaffected. The same behavior is seen in the deconvolution
of both the Pf$\gamma$ and the Br$\alpha$ images, but the artifacts appear
earlier for the slightly overexposed Br$\alpha$ image.

Indeed, the presence of a gap of this size does not contradict
our other knowledge of Eta Car. The deconvolved Pf$\gamma$ and
Br$\alpha$ images reach a spatial resolution of about 60~mas. This
scale is particularly interesting since it is close to the
expected radius where dust sublimates.

Using the approximation of a black-body equilibrium temperature,
we end-up with the formula presented by Smith et al. 2003a (Eq.3):
\begin{equation}
T_{\rm dust} \simeq 13100 \times R^{-1/2}_{AU} K
\end{equation}

We see that a dust temperature of about 400 K is expected at 900
AU or 0$\farcs$4  (if shielding from closer material is not
taken into account). Taking a typical sublimation temperature of
about 1000 K, we find that this radius is expected at about 150
AU, i.e. 70~mas from the central star. Of course, the dust
sublimation radius is only an indicative distance; the amount
of dust will steadily decrease as the temperature increases beyond
the sublimation point of each dust species. In the following
section we provide some indications on the dust composition in the
nebula.

This gap between the central source and the Weigelt complex may
have another explanation. Dorland et al. (2004) and Smith et al.
(2004a) have presented evidences that the Weigelt blobs C and D
were created in an outburst, either in 1941, or in
1890. If no dust has formed in the equatorial plane since then,
then the gap is a natural consequence of the proper motion of the
Weigelt blobs and not related to the temperature near the central
star.

The question of the vertical extent of the Weigelt complex is also
a difficult one. Hillier and Allen (1992 ) argued that the central
source is occulted by dust, while the Weigelt blobs suffer much
less circumstellar extinction. The location of this extinction is
somewhat uncertain. Why should the star be occulted, but not the
Weigelt blobs? Moreover the central source has brightened
appreciably over the last decade. At V it is now a factor of 3
brighter (see Davidson et al 1999, and recently Martin et al.
2004). The simplest interpretation is that the extinction is
decreased, and hence dust is evaporating, leading to a larger void
region around the star. Interestingly, van Genderen \& Sterken
have shown that this brightening occurred in a relatively short
time after the 1998.0 spectroscopic event attributed to the
periastron passage a hot companion (see also
Sec.~\ref{sec:binarity}).

In the deconvolved images (see Fig.~\ref{fig:NACOdeconv}) a large
part of the nebulosity has disappeared in the treatment, but in
Fig.~\ref{fig:color} we can see that the star is somewhat
embedded. In L$'$ the dust becomes more and more optically thin
and the regions towards the line of sight are difficult to detect.
This can be done only by a careful study involving several
filters, to map the extinction and evaluate the amount of
scattering. This study could be performed with carefully
calibrated NACO images but this implies dedicated observations
which will be postponed to a future study.

\section{Dust composition and temperature}

\label{sec:dust}
\begin{figure}
  \begin{center}
      \includegraphics[height=6.cm]{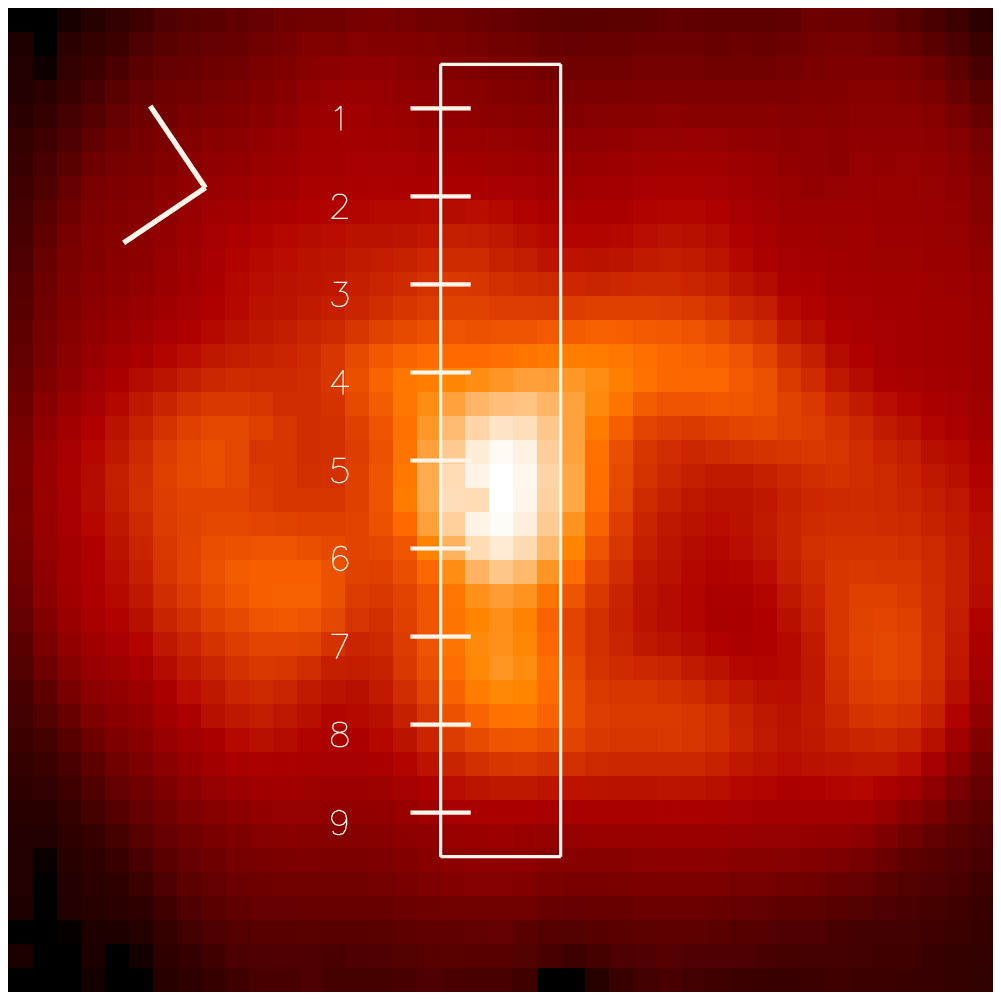}
  \end{center}
  \caption[]{MIDI 8.7~$\mu$m acquisition image. The
north points towards the upper-left side and east the lower-left
and the position of the slit is indicated. The slit is 0$\farcs$6
wide and 3$\arcsec$ long. The numbers indicate the central
positions of the apertures described in Table~\ref{tab:dust}.
\label{fig:slit}}
\end{figure}

\begin{figure*}
  \begin{center}
      \includegraphics[width=18.cm]{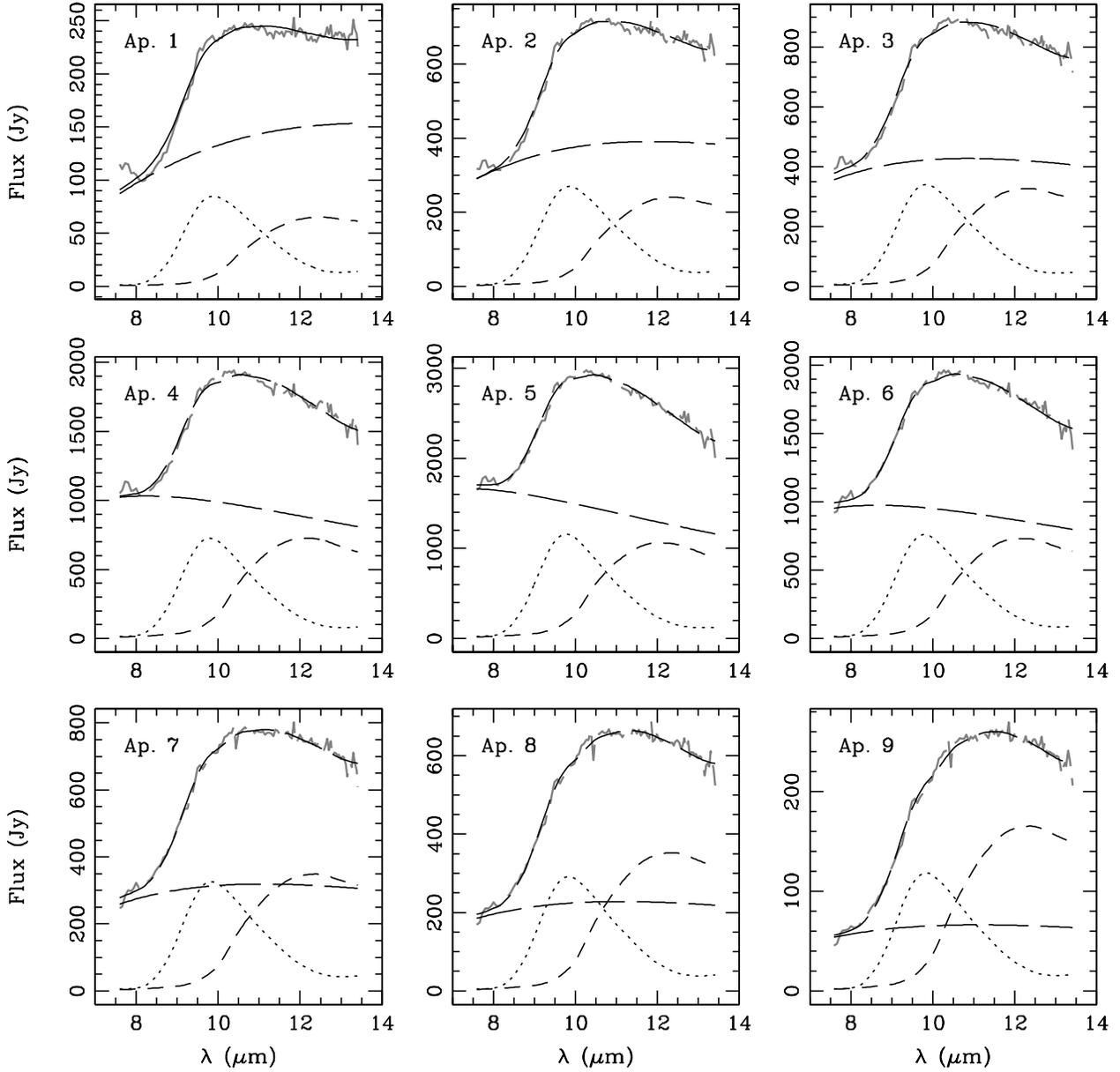}
  \end{center}
  \caption[]{\label{fig:specmodel}Spatially resolved MIDI spectra expressed in
  Jansky (grey lines) together with the spectra of the best fit models (solid
  line). The dotted line shows the olivine contribution, the dashed line the
  corundum contribution and the long-dashed shows the continuum emission. The
  spectra are extracted from the north-west (upper left panel) to the
  south-east (lower right panel). The slit is aligned to the nebula. The 9
  spectra are spaced by 0$\farcs$4, the maximum flux in the MIR is in aperture
  5 while the star is located in aperture 6. The SE clump and the southern arcs
  spectra are in apertures 7, 8 and 9. }
\end{figure*}

\begin{table}[h]
 \caption{\label{tab:dust} Positions and beam size of the apertures used to
 extract the spatially resolved spectra along the main axis of the nebula,
 increasing numbers from north to south. The three last columns report the
 result of the best fit to the spectra by using thermal emission with
 temperature T$_{bb}$ and opacities computed for various dust species.}
 \vspace{0.3cm}
\begin{center}
 \begin{tabular}{crr|ccc}
 \hline
  \hline
{\small Ap.} & \multicolumn{1}{c}{{\small Shift} } & {\small FWHM} & {\small T$_{bb}$} & {\small  Silic.} & {\small Al$_2$O$_3$}\\
        & \multicolumn{1}{c}{(mas)} &\multicolumn{1}{c}{(mas)} & (K) & (\%) & (\%)\\
 \hline
%\multicolumn{4}{c}{23/24-02-2003, JD=2452695, $\Phi=0.91$, B=?, $\Theta=?$}\\
%0    &  -1960 &     365 & \\

1  &  -1568 &     350  & 310$\pm$50&75$\pm$15 & 25$\pm$15 \\

2 &-1176  &    330 & 390$\pm$50& 65$\pm$10 & 35$\pm$10 \\

3 &-784   &   200 & 460$\pm$50& 60$\pm$10 & 40$\pm$10  \\

4 &-392 &200 & 600$\pm$70 & 55$\pm$5 & 45$\pm$5 \\

5 & 0 &200 & 720$\pm$80 & 55$\pm$5 & 45$\pm$5 \\

6 &392 &     200& 570$\pm$5 & 55$\pm$5 & 45$\pm$5 \\

7 &784 &200 & 440$\pm$50 & 55$\pm$5 & 45$\pm$5 \\

8 &1176 &330 &440$\pm$50 & 50$\pm$5 & 50$\pm$5 \\

9 &1568   &   350 &480$\pm$60 & 45$\pm$5 & 55$\pm$5 \\
\hline
  \end{tabular}
\end{center}
 \end{table}

In this section we will concentrate on the interpretation of the 9
single-dish N band MIDI spectra. The N band spectra of Eta Car are
characterized by a strong, smooth feature around 10.5~$\mu$m. The
feature has an unusually broad wing at the long wavelength side.
The 9 MIDI spectra, shown in Fig.~10, display a change in the
emission feature as a function of position in the nebulae. From
the north to the south the peak position is shifted from 10.5 to
11.5~$\mu$m.

In order to study the mineralogy of the dust we made an attempt to
fit the 10~$\mu$m spectra. The spectrum in the 10~$\mu$m region is
dominated by thermal emission from warm ($T>250$\,K), small
($a<2\,\mu$m) dust grains. Colder grains will emit most radiation
at longer wavelengths while big dust grains will contribute mainly
to the continuum which makes the determination of their mineralogy
difficult. We use here a very simple model consisting of a single
blackbody source function with two different dust species,
amorphous olivine (MgFeSiO$_4$) and corundum (Al$_2$O$_3$). We
also add continuum emission with the same temperature. The choice
of the dust components will be discussed below. We take a single
grain size of $0.1\,\mu$m. Using a more complicated source
function involving a distribution of temperatures or by including
more dust species or grain sizes did not improve the fit
significantly. By including more dust species we find that some
trace of crystalline olivine might be present but with an
abundance less than 5\%. In order to calculate the emission
efficiencies of the dust grains, we have to assume a shape of the
dust grains. The choice of the particle shape model can be crucial
in obtaining reliable results. However, since both dust components
used here have a rather smooth behavior we restrict ourselves to
simple, frequently used methods to calculate the emissivities. The
best fitting results were obtained if we take the amorphous
olivine grains to be homogeneous and spherical. For the corundum
grains we had to use a so-called continuous distribution of
ellipsoids (CDE) (Bohren \& Huffman 1983) to reproduce the
observations. The abundances are obtained by using a standard
linear least square fitting procedure. This simple model gives us
an indication of the composition of the small, warm dust
component, and, using the observed MIDI spectra, provides a
quantitative way to study the spatial variation in the dust
composition.

We include amorphous olivine in our fitting procedure because it
is one of the most abundant dust species in circumstellar,
cometary and interstellar dust. Corundum is expected to be the
first species to condense at very high temperatures (1700\,K) and
preferably at high densities (corresponding to a pressure of
$10^{-3}$\,atm, see e.g. Tielens 1990). Mitchell \& Robinson
(1978) showed that a substantial amount of corundum is needed in
order to fit the large aperture 10~$\mu$m spectrum of Eta Car.
The spectrum of Eta Car obtained by ISO, presented by Morris et al
(1999), provides additional evidence for the presence of
non-silicate dust. The reason is that if all of the emission
around 10~$\mu$m would be caused by silicates, this would generate
an appreciable 18~$\mu$m emission feature, which is not observed.
Moreover, the broad red wing of the 10~$\mu$m feature extends to
15.5~$\mu$m, which is not characteristic for silicate emission.
Another argument that the 10~$\mu$m emission should contain a
significant component of non-silicate dust is the lack of a clear
detection of crystalline silicates as seen in some other LBVs
(Waters et al, 1997). As it is hard to explain that in Eta Car
only amorphous silicates would form, the lack of observable
silicate crystals suggests that the dust material is not
completely dominated by silicates.  Finally, we note that corundum
is also expected to be present in the ambient environments of
these types of stars. In recent studies of AGB stars for instance,
abundances of corundum of about 10-30\% are reported (see the
extensive discussion in Maldoni et al. 2004). Moreover, the
appearance and disappearance of the corundum signature in the
variable spectra of pulsating OH/IR stars is interpreted as
evidence for dust formation (Maldoni et al. 2004). The above
arguments strongly favor a grain component in addition to
silicates. As corundum can naturally explain the red-wing of the
10~$\mu$m feature, we adopt this species as the extra component.

It is worth discussing that Mitchell \& Robinson (1986) discarded
the possibility of corundum as a dust component. They did so
because \emph{i}) corundum is not expected to survive as a
separate grain component as it would act as nucleation centers for
the later precipitation of silicates, and \emph{ii}) the required
amount of aluminum suggested an unrealistic overabundance compared
to the solar value. Mitchell \& Robinson fitted the broad
long-wavelength shoulder of the 10~$\mu$m feature (which we
attribute to corundum, see below) by emission from large
(2~$\mu$m) amorphous silicate grains that display a very broad
10~$\mu$m feature. For their calculations they use the refractive
indices of 'astronomical silicate' as derived by Draine \& Lee
(1984). However, calculations for large amorphous olivine grains
using laboratory measurements of the refractive indices, by for
example Dorschner et al (1995), show a feature that is less
broadened and is incompatible with the observed red wing of the
10~$\mu$m MIDI spectra.  Concerning their first argument, one
should keep in mind that the Eta Car nebula is expected to be a
CNO processed medium (Davidson et al. 1986, Waters et al. 1997,
Smith et al. 2004b). This could lead to a condensation sequence
that is different from that which occurs in other stars favoring
the creation of corundum. In Eta Car, the oxygen that remains
after the CO molecule formation could be so modest relative to the
amount of metals present that part of the material is only able to
form simple oxides, such as Al$_2$O$_3$. (Also, one may expect a
chemistry driven by remaining metals, notably sulphur, forming
species such as MgS.) An alternative explanation for the presence
of corundum may be that the gas density at the location of dust
formation is so low that it is not possible to complete all of the
condensation sequence leading to silicate dust, i.e. the
condensation reactions freeze out. Concerning the second argument
of Mitchell \& Robinson, note that both possibilities discussed
above may, at least in principle, explain the apparent over
abundance of \emph{solid state} aluminum relative to silicon.

The results of the analysis as described above are summarized in
table 4, the 9 spectra are shown in Fig.~10 together with the best
fit model. Also shown are the contributions from the various
components. We see that when we go down from north to south the
abundance of corundum is increased. This is consistent with the
observed shift of the feature towards longer wavelengths when
going from north to south.

It should be noted that the derived abundances are subject to a
correct estimate of the continuum contribution. In our simple
model the contribution from cold dust grains to the continuum
emission is not taken into account. Including this in a more
complicated model might cause changes in the derived dust
composition. To test the effect of grain size on the derived
abundances we have performed calculations in which large amorphous
olivine grains were added. This reduced the derived abundance of
corundum in all fits by $\sim 5$\%, i.e. the trend in the
compositional gradient is not significantly effected.

The evolution seen in the spectra is indirect evidence that dust
is continuously created in the butterfly nebula or at least that
the geometry of the butterfly nebula strongly influences the
chemical composition of its dust content. The aperture 1 spectrum
is dominated by emission from olivine grains. This aperture is
pointed towards the region where the Weigelt complex ends and
probably encounters, in the equatorial plane, the walls of the
polar lobes. The dust in this region is efficiently shielded from
the light of the central object. The aperture 7-9 spectra are
dominated by the emission from aluminum oxide grains.  A possible
explanation for this might be that the condensation reactions
freeze out. In this scenario the difference in dust composition
between the Weigelt complex and the SE clump reflect a different
formation process; the equatorial dust being formed preferably
during outbursts which provide dense enough regions to complete
the condensation process whereas the dust formed in the rims of
the butterfly nebula is continuously processed but the reactions
are quickly freezed out. An argument against this scenario however
is that the impact of the wind through the rims should provide a
density discontinuity large enough to provide the conditions of
dust silicate formation. Spectra taken {\it beyond} the rims of
the butterfly nebula are needed to constrain the condensation
sequence and begin a study of the chemical map of the dust within
the full nebula.

\section{Correlated flux}
\label{sec:correlated}

\subsection{Location of the fringes}
\label{sec:spatfringes} 
\begin{figure*}
  \begin{center}
      \includegraphics[height=8.cm]{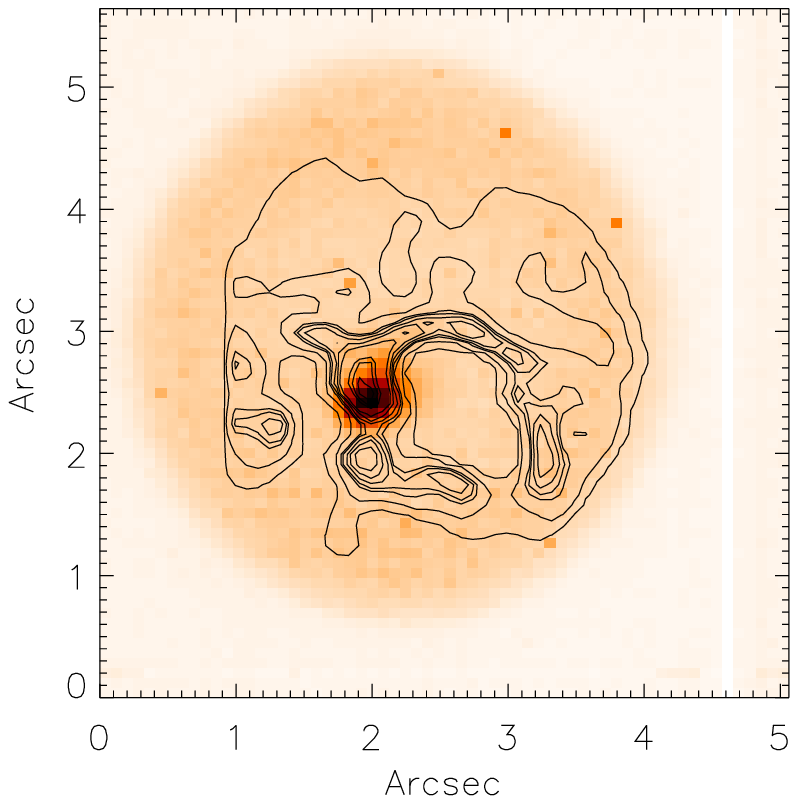}
  \includegraphics[height=8.cm]{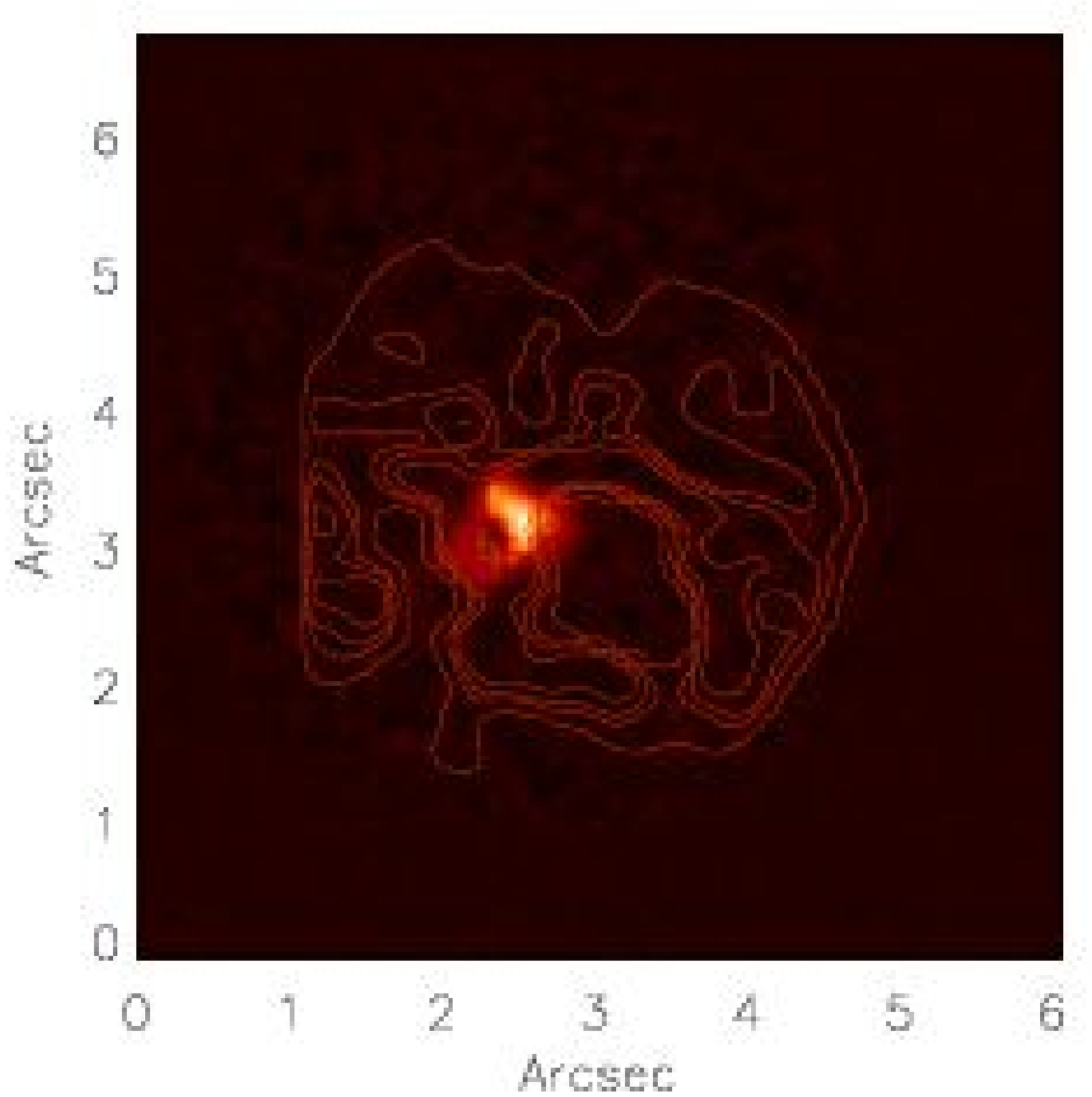}
  \end{center}
  \caption[]{Left, the figure shows the rms of the fluctuations within the
MIDI FOV. The external regions are dominated by the detector noise
and the internal regions by the tunnel and sky background
fluctuations. The signal from the fringes is strong and centered
on the position of the star as seen in the deconvolved acquisition
image at 8.7~$\mu$m. The contour plot represents the contours of
the deconvolved MIDI 8.7 $\mu$m acquisition image. Right, the
noise pattern  and the fringe pattern from a calibrator have been
subtracted from the previous figure in order to show the extended
fringe signal. The orientation is the same as in
Fig.~\ref{fig:slit}. \label{fig:spatfringes}}
\end{figure*}

The spatial distribution of the fringes
detected by MIDI with the 8.7~$\mu$m filter is shown in
Fig.~\ref{fig:spatfringes}. We have checked that the peak of the
fringes is localized at the position of the star
itself\footnote{by comparing the position of the fringes with the
position of the emerging peak in the deconvolution image
(Fig.~\ref{fig:DeconvMIDI}.} but an extended halo is also visible
in the Weigelt complex about 0$\farcs$4-0$\farcs$6 northwest from
the star. This is the confirmation that highly compressed material
which emits strongly at 8.7~$\mu$m exists in this region.

The fluctuations from the fringes at the location of the Weigelt
blobs are definitely more extended than a single PSF FWHM at
8.7~$\mu$m (220~mas). They coincide roughly with the location of
the blob C' observed by NACO but are more extended owing to the
larger PSF of the 8m telescope at this wavelength. This implies that
in the equatorial Weigelt region a fraction of the dust is
embedded in clumps with a typical size smaller than 10-20~mas
(25-50~AU) within a total extent of about 1000 AU. Nevertheless, this
correlated flux represents only a few percent of the total flux at
these locations. It must be pointed out that only a few scans with
fringes have been recorded during this commissioning measurement
and the lowest detectable fringe signal visible in
Fig.~\ref{fig:spatfringes} is about 20~Jy. In June, the
measurements performed in dispersed mode (following section)
represents more than 200 scans. MIDI has been able to record
fringes further out from the central source with a sensitivity
reaching about 5~Jy. From this result we are confident that the
spatial distribution of the correlated flux can be studied in
the future at distances larger than 0$\farcs$5 from the central
object.

\subsection{Correlated spectra}
\label{sec:specfringes}
\begin{figure*}
  \begin{center}
\includegraphics[height=10.cm]{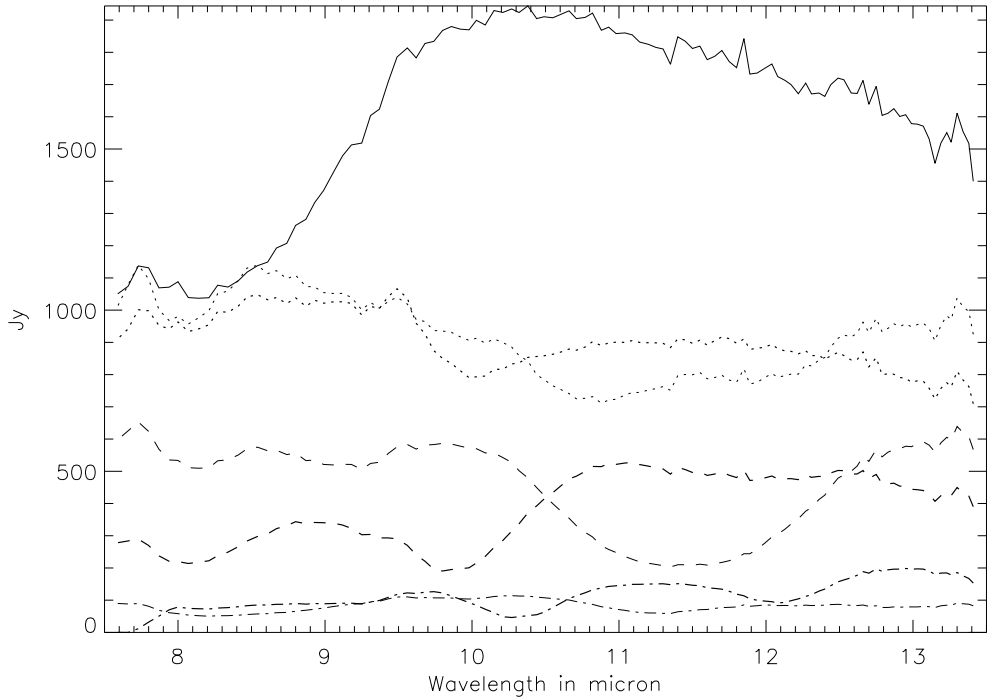}
  \end{center}
 % \caption[]{MIDI correlated flux measured with a 78m projected
 % baseline. The solid line denotes the photometric flux as
 % extracted using a mask centered on the star. The dotted line
 % represents the correlated flux measured with the same mask. The
 % scale has been multiplied by 10. The dashed and dashed-dotted lines
 % are represent the correlated flux extracted with different masks
 % located at +0.4$\farcs$ (dashed) and -0.4$\farcs$ (dashed-dotted) referred to
 % the nebula main axis (positive to the north). The stars indicate
 % the expected correlated fluxes {\it multiplied by 4 and not 10, i.e. divided by 2.5}.
 % The crosses indicate the same point with a scaling of 3.

 \caption[]{MIDI correlated flux measured with a 74m (PA=62$^\circ$) and a 78m (PA=56$^\circ$) projected
 baseline. The solid line denotes the photometric flux as
  extracted using the mask centered on the star (aperture 6) which can be seen in Fig.~\ref{fig:specmodel}.
  The dotted lines represent the correlated flux measured with the aperture 6.
  scale has been multiplied by 10. The dashed and dashed-dotted lines
  represent the correlated flux extracted with aperture 5 (dashed)
  and 7 (dashed-dotted).
  %The stars indicate
  %the expected correlated fluxes from Hillier et al. model (2001) {\it multiplied by 4 and not 10, i.e. divided by 2.5}.
\label{fig:specfringes}}
\end{figure*}

In Fig.~\ref{fig:specfringes}, we show the correlated flux of
three central masks around the star (apertures 5 to 7). For
comparison, the photometric flux of aperture 6 is shown (this
aperture contains the star, the maximum flux being in aperture 5).
The correlated fluxes measured by MIDI are 98$\pm$47~Jy,
87$\pm$33~Jy and 82$\pm$41~Jy, respectively. The errors have been
estimated from the variance of the measurements using several
calibrators and by varying slightly the parameters of the
apertures. The fringes have been recorded using the same slit as
used for the photometry. Unfortunately, this slit is about two
PSFs wide and the signal from the star has been mixed up with the
signal from the dust situated perpendicular to the axis, i.e. at
PA$\simeq$60$^\circ$.

The baseline is roughly perpendicular to the main axis of the
nebula and van Boekel et al. (2001) have reported that the star is
prolate. This means that the baselines were oriented perpendicular
to the main stellar axis, where the star is smaller, corresponding
to a maximum correlated flux. Hence, our measurement can be
considered as an upper limit of the correlated flux observable
from the star.

We compared these measurements with the model presented in Hillier
et al. (2001). For that purpose, we used three flux distributions
from the model at 8, 10 and 13~$\mu$m, of respectively 332, 287
and 241~Jy which can be approximated by a 2D gaussian with a FWHM
equal to 6.4, 6.8 and 8.2~mas. With the UT1-UT3 projected baseline
of 78m, we can compute the expected correlated flux by performing
the Fourier transform of the flux distribution from the
(spherical) models. The visibility for the theoretical star at 8,
10 and 13~$\mu$m is 0.54, 0.59 and 0.65 respectively. This
corresponds to correlated fluxes of 180, 169 and 156~Jy. These
fluxes are larger than those observed by a factor of 2. Moreover,
the correlated flux measured at the location of the central star
includes also a non-negligible contribution from the dust.

The correlated flux from the Weigelt region is dominated by the
brightest clumps. Due to the complexity of their spatial
distribution, the curves of correlated flux present an oscillating
behavior which is very dependant on the projected baseline length
and direction. This is particularly visible in the correlated flux
spectra extracted from aperture 5 (dashed line in
Fig.~\ref{fig:specfringes}). In particular, the frequency and the
stability of this oscillation over the N band suggest that a few
clumps separated by 0$\farcs$05-0$\farcs$1 dominate the correlated
flux of apertures 5 and 7. The correlated flux extracted from
aperture 6 (which contains the star) is larger and the oscillation
much lower suggesting that the dust contribution is relatively
low, about 10-20~Jy, compared to the stellar flux. We are left
with a stellar flux of about 70-90~Jy at 8~$\mu$m and about
50-70~Jy between 10 and 13~$\mu$m.

The correlated fluxes represent about 50~Jy in the location of the
Weigelt complex and only 5-10~Jy in the south. If we compare these
correlated fluxes with the total measured fluxes, the visibility
and hence the clumping factor are larger at the location of the
Weigelt complex (more than 3\% visibility) than at the SE clump
(less than 2\%) though this difference is within the MIDI error
bars. We are quite confident that even the smallest correlated
fluxes reported here are real. No correlated flux can be detected
at the northern edge of the slit. At this location, the flux from
the nebula is still well above the detection limit of MIDI, which
is of the order of one Jansky for faint fluxes. Moreover, MIDI has
observed some bright overresolved sources without showing spurious
fringe detection. For instance, no fringe signal was detectable
for the bright source OH~26.5+0.6, an OH/IR star with a N band
flux at the time of our observations of $\approx$~650~Jy (Chesneau
et al. 2004).

\section{Discussion}
\label{sec:discussion}

%\subsection{2.2$\mu$m-13.5$\mu$m SED of the central object}
\subsection{Inferences for the observations of the central star}

As noted in section 5.2, the IR fluxes deduced from the present
observations, and by van Boekel et al 2003, are a factor of 2 to 3
smaller than those predicted by the model. Below we discuss the
possible cause of these discrepancies.

The first cause we examine is the correction for reddening. From
the inferred dust temperatures, and from the JHKL variability
observations of Whitelock et al. (2004), we can infer that the K
band flux of Eta is dominated by the central source, and by
scattering. Feast, Whitelock and Marang (2001) give an IR
magnitude for Eta Car of around 0.4 to 0.5. This, and the
stellar K magnitude of 1.2 derived by van Boekel, implies that half
the starlight is scattered. Thus there is considerable extinction
at K, and this extinction could easily explain the difference
between the van Boekel K flux and the model K flux. However at
10~$\mu$m, the extinction will be lower, and probably cannot
explain the discrepancy. Moreover, a variable free-free emission
seems to be also an important flux contribution in K band which is
also contaminated by the emission from the Br$\gamma$ line
(Whitelock et al. 2004). The complexity of the K band is such that
the constraints provided in N-band should be more reliable. Thus we
must look to the modelling for an explanation of the discrepancy.

%To model Eta Carinae we assume that the primary component is the
%most luminous, and that it is the primary that dominates the
%mass-loss rate from the system, and controls the observed stellar
%spectrum. X-ray modeling suggests that while the secondary wind
%may have a higher terminal velocity, it has a substantially lower
%mass-loss rate (ref).

%For the modeling we have an excellent estimate of the distance
%(2.3 \pm 0.2 kpc; e.g. Allen and Hillier 1993), and the system
%luminosity  (> 4.1 \times 10^5 Lsun) from IR measurements (Smith e
%al. 2003). Normally these would allow a very precise estimate of
%the stellar parameters, but unfortunately, as will be discussed
%later, Eta Car is not that simple.

%Using HST spectra obtained in March 1998, and assuming
%N(He)/N(H)=0.2, Hillier et al (2001) derived a mass-loss rate of
%1.0D-03 Msun/yr with a filling factor of 0.1. The mass loss rate
%is primarily derived from the equivaent widths of the Balmer
%lines, while the filling factor is constrained by the strength of
%the electron scattering wings. The VLT observations of van Boekel
%et al (2003) suggest Mdot=1.6D-03Msun/yr with f=0.225. However,
%with this value the electron scattering wings appear to be
%somewhat too strong. The flux distributions of the two models are
%very similar.

There are some major difficulties associated with modelling of the
optical/UV spectra of Eta Car.
\begin{enumerate}
\item We cannot compare the model fluxes with those observed since
the reddening and reddening law are uncertain, and have to
themselves be derived from the observations. In addition, the
amount of circumstellar reddening is probably variable.

\item There is evidence for a possible wind asymmetry. This might
explain  explain why Hillier et al. severely overpredicted the
strength of the P Cygni absorption lines seen in optical spectra.
Direct evidence for an asymmetry comes from the variable terminal
velocity derived from the scattered H$\alpha$ profiles (Smith et
al. 2003b), and from VLT measurements (van Boekel et al 2003).

\item The companion star could be a substantial
source of ionizing photons which could also affect the symmetry of
the wind. The large impact of the orbital cycle on the
near-infrared photometry is a strong argument for it (Whitelock et
al. 2004).

\item The spectrum of the primary is intrinsically variable, and part of the variability is
probably not attributable to a companion. In particular during
2002, and leading up to the event in 2003, the H Balmer lines were
up to a factor of 2 weaker compared to the previous cycle. The
behavior of the radio emission is also different from the last
cycle (Duncan \& White 2003).

\end{enumerate}

Given all these difficulties it is not surprising that the agreement
of model and observations is not perfect. The match of the
model with a large part of the spectrum from the central object
is already a success. However it is worth examining in more
details possible causes of the discrepancies. We consider two
possible causes: variability and wind asymmetry.

Since the IR flux will originate where the wind is at a substantial
fraction of the terminal velocity, we can use the mass-loss rate
formula of Wright \& Barlow (1975) to estimate the scaling of the
IR flux with mass-loss rate. In particular, $ \dot{M} \propto
 S^{0.75}$. Thus a factor of 2(3) reduction in the IR flux
corresponds to a change in the mass-loss rate of a factor of
1.7(2.3). Using HST spectra obtained in March 1998, and
assuming\footnote{In the modelling there is a strong coupling
between the derived H/He abundance ratio and the mass-loss rate.}
N(He)/N(H)=0.2, Hillier et al (2001) derived a mass-loss rate of
$1 \times 10^{-3} M_{\odot}$/yr with a filling factor of 0.1. The
mass loss rate is primarily derived from the equivalent widths of
the Balmer lines, while the filling factor is constrained by the
strength of the electron scattering wings. The VLT observations of
van Boekel et al (2003) suggest $\dot{M}$=$1.6 \times
10^{-3}M_\odot$/yr with f=0.225. However, with this value, the
electron scattering wings appear to be somewhat too strong. The
flux distributions of the two models are very similar. It is worth
mentioning that the HST spectrum of March 1998 has been taken at
an orbital phase very close to the one of MIDI measurements. In
contrary, the VINCI measurements, have been carried out in the
first half of 2002, i.e. at a very different part of the cycle. It
is possible in this context that the mass-loss rate and geometry
were strongly affected (Smith et al. 2003b).

As noted previously the H$\alpha$ profiles have changed, and their
weakening could be interpreted as a reduction in mass-loss rate.
Since the H$\alpha$ and 10~$\mu$m emission come from a similar
volume (the H$\alpha$ volume is slightly larger) it is not
surprising that a reduction in IR flux accompanies the reduction
in H$\alpha$ flux. An alternative scenario for the variability is
that the flux that maintains the ionization of the wind has been
reduced. The existence of strong FeII emission lines, the radio
variability observations (e.g., Duncan and White 2003), and the
models show that H recombines in the outer envelope.

A second explanation is a wind asymmetry. A wind asymmetry will
certainly bias our derived mass-loss rates. However a wind
asymmetry will generally have substantially less influence on the
K-10~$\mu$m color, simply because the stellar fluxes at both IR
wavelengths are produced by free-free processes, and hence are
affected in the same way.

Clearly repeated quasi-simultaneous observations, at 2 and
10~$\mu$m, are very important to ascertain the consistency of the
model constraints. This will be possible soon with the advent of
(quasi)-simultaneous observations of Eta Car with MIDI in the
near-IR interferometer AMBER (Petrov et al. 2003)

\subsection{Geometry of the dusty inner nebula}
The Homonculus shape has been discussed by many authors, but it is
only recently that NIR spectroscopy allowed unambiguous tracing of
the shape and orientation of the dense neutral gas and dust
through the observation of the H$_2$ emission (Smith, 2002b). In
particular, Smith demonstrates that near the equator the walls of
the bipolar lobes do not converge towards the central star (see
his Fig.7). A simple extrapolation of the H$_2$ data indicates
that the connection of the lobes with the equatorial plane takes
place at about 2000-4000 AU, i.e. about 1$\farcs$3 from the
central object. This distance is compatible with the projected
mean position of the rims of the dusty inner nebula seen in the
NACO and MIDI images. This naturally led Smith et al. (2002a,
2003a) to suggest that the complex structures seen in their images
indeed lie close to the equatorial plane like it has been proven
for the Weigelt complex. They tentatively explain the complicated
shape of this equatorial structure in the frame of a preexisting
torus disrupted during the great eruption of 1840 or by the post
eruption stellar wind. The Weigelt complex, which also lies on the
equatorial plane would have been ejected later on, in the second
eruption of 1890.

It is indeed very difficult from images only, and without any
kinematic information from the structures to get a 3D view of the
object. Within this context any model would be highly conjectural,
yet we propose in this section some arguments suggesting another
point of view. The images show a highly structured butterfly shape
which is well delimited by bright rims. In particular, we have
shown that the SE clump is a warm head of a protruding region
linking the SE and Southern arcs which exhibits a large amount of
corundum. The SE clump seems to be closely aligned with the polar
axis of the star and the bipolar nebula. This is for us an
indication that this structure could directly face the fast and
dense wind of Eta Car, and therefore not lie in the equatorial
plane. The rims of the dusty inner nebulae seems also to share a
similar axis. In particular, the Western and Northern arcs (see in
Fig.~\ref{fig:sketch}) appear to converge to a point lying within
or behind the Weigelt complex. It must be pointed out that this
position is rather symmetrical to the position of the SE clump.
The butterfly shape itself is suggested by two other protruding
regions, namely the NE and the SW clumps (Fig.~\ref{fig:sketch}).
Such a symmetry is potentially highly informative on the physical
processes acting close to the star. Could it be that these
structures are the sky projection of 3D optically thin geometry?
Such an interpretation is at the moment premature. It should be of
greatest interest to measure the radial velocity of the rims, but
the combination of spectral and spatial resolution required is
difficult to attain \footnote{The NACO spectral resolution in
spectroscopic mode being limited to R$\approx$1500, i.e 200 km
s$^{-1}$}.

A complex relationship must exist between the IR Butterfly nebula
and the Little Homunculus discovered by Ishibashi et al. (2003)
with the HST which is supposed to be also a consequence of the
eruption of 1890. This structure is seen in emission lines at
visual wavelengths, while the IR images are dominated by dust
emission which makes it difficult to easily compare the two
geometries. However the similarity of their spatial extensions
(about 2$\arcsec$) probably points to a common origin of the
structures. Ishibashi et al. (2003) showed that the polar caps of
the Little Homunculus are expanding outward at about
300~km~s$^{-1}$. The present polar wind is much faster, of the
order of 1000~km~s$^{-1}$ and it carries a high flux of mass
(Smith et al. 2003b, see in particular their Fig.7). Dwarkadas \&
Owocki (2002) predict a mass-flux difference of a factor of about
five in the polar and equatorial direction. At this speed the wind
ejected about 40~yr ago should have impacted this preexisting slow
motion structure supposedly ejected in 1890. Of course we assume
that the latitudinal dependance detected in 2003 was already
present by that time. We suspect that the conditions for dust
formation could be phenomenologically equivalent to the ones
encountered in the dusty WR+O binary as proposed recently by Smith
et al. (2004a). The rims of the butterfly nebula is probably a
place where a strong density gradient is combined with high
temperatures. The fast current wind of Eta Car may impact strongly
these rims providing the conditions for an efficient dust
formation.

Any slow dense material in the vicinity of Eta Car (i.e. within
2$\arcsec$) has to face three spatially localized regimes of wind.
The polar regions of the inner nebula are facing a dense and fast
wind, the intermediate-latitude regions experience fast but
probably less dense wind, and the equatorial regions receive an
equatorial wind with considerably less kinetic energy. Moreover
there is a considerable shielding close the the equatorial plane
in the direction of the Weigelt complex. It is well established
that this zone (which contains the so-called 'strontium region')
presents fairly low excitation condition compatible with an
efficient dust processing (Hartmann et al. 2004) but it is also
relatively devoid of dust compared to other parts of the
Homunculus nebula. From the previous considerations, we expect a
latitudinal modulation of the survival probability of any dense
dusty structure in the vicinity of Eta Car.

Finally, the consequences of the binarity of Eta Car are probably
large and we discuss some potential consequences of the wind-wind
collision on the dust lying close to the equatorial plane.

\subsection{Effect of the binary orbit on the equatorial ejecta}
\label{sec:binarity} The Weigelt complex has been extensively
discussed in many papers (Weigel \& Ebersberger 1986, Hoffman \&
Weigelt, 1988, Weigelt et al. 1995, 1996) and it is now well
established that it lies at least close the equatorial plane
(Davidson et al. 1996, Smith et al. 2004a, 2004d). As it has been
said previously, in the NACO images the Weigelt complex embeds the
star and there is no apparent disconnection between this structure
and and so-called SE filament, which shares the same apparent axis
of symmetry. In this section we tentatively attribute the shape
and mass content of the SE filament to the effect of a wind-wind
collision with a secondary star during and after the outbursts.

The SE filament could be the physical counterpart of the Weigelt
complex, on the partly obscured receding part of the equatorial
plane. This might explain the flux difference between the two
regions although a differing illumination from the central star
may also help. However, a more straightforward explanation is that
the amount of material is simply much weaker than the one seen in
the Weigelt complex. The absence of dust emission in the
North-East must also be integrated to a global interpretation.

The global effects of the 5.52 cycle on the X ray (Corcoran et al.
2001), the optical (Damineli, 1996, Damineli et al. 2000, Smith et
al. 2000a, van Genderen et al. 2001, Martin et al. 2004) and the
NIR (Smith et al. 2000b, Whitelock, 2004) flux are now better
understood, but the secondary characteristics and its orbital
parameters are still unconstrained. Two facts appear unavoidable:
the period of the orbit is rigorously established and the
eccentricity of the system must be large.

If we place the orbit of the binary such that the periastron is
located between the Weigelt complex and the star, and the apastron
in the North-East we can interpret the apparent equatorial
structures by a competing effect between the winds and radiative
fluxes of the primary and the secondary. When the secondary is at
periastron, its cone wind dominated by a fast and diffuse gas
passes the Weigelt complex relatively rapidly, limiting the amount
of dust destroyed at each passage and explaining the remanence of
such a large amount of material close to the star. The binary
model could also help explaining the formation of an equatorial
ejection of large and dense clumps at low velocities during the
outburst of 1890. On the other hand, the dust material situated in
the North-East could have been efficiently cleaned out during the
18 orbits since the 1890's outburst by the slower passage of the
secondary wind cone. Finally, dust may still form in the
equatorial plane refuelling the SE filament and the Weigelt
complex at each orbit like seen in pinwheel nebulae (Tuthill et
al. 1999). The dichotomy between the North-East empty region and
the SE filament is easily explained in the frame of a pinwheel
nebula by the tilt angle of the cone of the secondary wind
compared to the secondary orbital motion.

This hypothesis can be tested by a monitoring of the UV emission
in this region (part of the "purple Haze"). Recently, Smith et al.
(2004a, 2004b) presented such a monitoring of the emission from
the "purple Haze" and proposed a model of binary orbit very
similar to the one what has been discussed above. However, they
place the periastron in the North-East of the primary star, while
we are tempted to put it in the South-West in order to explain the
large content and survivance of the Weigelt complex.

\subsection{Temporal variability}
Great care must be taken when placing the data presented here in
the frame of the 5.52-year cycle of Eta Car. We are dealing with
data of unprecedented spatial resolution and the variability
effects of the individually resolved features such as the
sublimation radius or the MIR emission of the Weigelt complex are
largely unknown. With the VLT and techniques like
Adaptive Optics and Long Baseline Interferometry the spatial
resolution is such that the expected time lag between a central
star event and its impact on our data is expected to be fairly
short. Flow time is given approximately by:
           $$t = 23.7\ (d /2.5kpc)\ (r / 1")\ (500 km.s^{-1} / v)\ yr$$
Thus for $d$=2.3 kpc, and $r$=0.1" the flow time is about 2.2 years.
The effect of a rapid change of the UV and visual luminosity of
the central object could affect the sublimation radius even
faster.

Both NACO images and MIDI data must consequently be situated in
the frame of the 5.52-year periodicity. Since accurate
photometry of the NACO images is difficult we rely more on a
monitoring of the geometry of Eta Car alone. Short time scale
variability ($\sim$ a month) can represent a real problem in
interpreting MIDI data at different projected baselines separated
by a large time lag. This question is of importance since it is
difficult to observe a large number of baselines during
a time interval smaller than the variability expected in mid-IR, which
corresponds to the time needed to form dust and move it away.

The NACO images have been recorded in December 2002 at phase 0.87,
but the most of the MIDI data have been recorded in June 2003 at
phase close to 1., just before the shell ejection event.
Whitelock, Marang, and Crause (2003) reported in June 2003 that
the anticipated fading of Eta Car at infrared wavelengths (JHKL)
started between June 19 (L-band) and June 24 (J-band), i.e. just
after the MIDI observations.

Large extinction effects are expected from putative shell ejection
from the primary star, which may be periodically triggered by a
companion. These shell ejections are implied by radio
observations (Feast et al. 2001). These ejections are expected to
offset the radius of dust sublimation which should be a great
indicator of dust formation and indirectly of the amount of
ionizing photons that can reach the different parts of the nebula.
Smith et al. (2000) have clearly demonstrated that the variability
detected by HST can be attributed either to a bolometric variation
of the star or, more probably, to grain destruction, which could
then explain a decrease in circumstellar extinction. A shift of the
dust forming regions may also affect the extinction.

Further observations with the same settings should provide more
information on the real performances of a monitoring of Eta Car by
the instruments NACO and MIDI.

%\begin{figure}
%  \begin{center}
%    \includegraphics[width=8cm]{sketch.eps}
%  \end{center}
%  \caption[]{\label{fig:sketchH2} Cartoon representing the geometry of the Butterfly dusty
%  nebula proposed in the article. This cartoon is inspired from the figure 7 in Smith 2002b. The
%  contours of the polar lobes (wide solid line) are traced by the H$_2$ emission detected in the NIR. The
%  contours of the butterfly nebula (solid line) are a representation of the size and orientation
%  of the structure with respect to the polar lobes as proposed in this article. The dotted arrow denotes the line of sight.}
%\end{figure}

\section{Conclusion}
\label{sec:conclusion} We have reported observations of Eta Car by
NACO and MIDI. We can draw the following conclusions:

\begin{itemize}
\item We have obtained near- and mid-IR images of the inner few arcsec
      of the core of Eta Car, using the VLT AO system NACO and the
      VLTI/MIDI interferometer, at an (unprecedented) spatial
      resolution of 60 and 140 milliarcsec respectively.
      The internal 3 arcsecs region of the dusty nebula has been observed in
      great detail.

\item The Weigelt complex of 'blobs' is seen in detail for
      the first time in the thermal infrared, and their position can
      be measured accurately. The IR counterparts of
      optical clumps C and D can be identified, whereas the clump B is
      not detected. These visible structures,
      dominated by scattering, may
      trace the walls of the dust clumps seen in the IR. A fainter structure,
      the SE filament may be the southern counterpart of the Weigelt complex.
\item The central object is seen by NACO to be separated from its
surroundings. A large area empty of dust of a typical radius of
0$\farcs$10-0$\farcs$15 (230-350~AU) is detected.

\item We have detected a chemical difference between the dust
located at the Weigelt complex and that located at the south-east
of the star. The aluminium oxide (corundum) seems to dominate the
spectrum of the SE clump. This molecule is one of the first dust
species to condense in a harsh environment.

\item These Adaptive Optics and Interferometric
observations allow the extraction of upper limits to the SED of the
central object from 3.7~$\mu$m to 13.5~$\mu$m.

\item We have discussed on the spatial geometry of the dusty nebula.
In particular, we have outlined that the bright dusty rims seem to
share the axis of symmetry of the Homunculus.

\item We hypothesis that the survival and the large mass of the Weigelt
blobs could be related to their location close to the periastron
of the system orbit, where the secondary star is strongly
extinguished by the primary dense wind. The counterpart of this
high density region is an almost empty one in the North-East and
the faint SE filament which could be the consequence of the
eccentric orbital motion of the secondary.
\end{itemize}

%__________________________________________________________________

\begin{acknowledgements}
Working on developing new techniques and new instruments is still
a great challenge which is always risky and not always rewarding.
The authors warmly thank the technical people of the MIDI team and
the ESO VLTI team working on Paranal observatory or in Garching.
They made possible the advent of MIDI whose performances can not
be dissociated from the ones of the impressive interferometric
infrastructure of the VLTI.
\end{acknowledgements}
%We thank J.A. Morse who agreed to make use of the HST images.

%We thank Dr. D.P.K. Banerjee for kindly provinding us the...

%\bibliographystyle{aa}   %>>>> makes bibtex use spiebib.bst
%\bibliography{SPIN}   %>>>> bibliography data in report.bib

\end{document}